\documentclass{article}

\usepackage{arxiv}
\usepackage{times}

\usepackage{soul}
\usepackage{url}
\usepackage[utf8]{inputenc}
\usepackage[small]{caption}
\usepackage{graphicx}
\usepackage{amsmath}
\usepackage{booktabs}
\usepackage{latexsym} 
\usepackage{amssymb}
\usepackage{amsthm}
\usepackage[group-separator={,}]{siunitx}
\usepackage{algorithm}
\usepackage[noend]{algpseudocode}
\usepackage{complexity}
\usepackage{multirow}
\usepackage{caption}
\usepackage{verbatim}
\usepackage{enumitem}
\usepackage{subfigure}
\usepackage{balance}
\usepackage{dblfloatfix}

\newtheorem{mytheorem}{Theorem}
\newtheorem{myproposition}{Proposition}
\newtheorem{mycorollary}{Corollary}
\newtheorem{mydefinition}{Definition}
\newtheorem{myexample}{Example}
\newtheorem{myremark}{Remark}

\newcommand{\myqed}{\mbox{$\diamond$}}

\newcolumntype{C}{>{\centering\arraybackslash}p{3cm}}
\DeclareMathOperator*{\argmax}{arg\,max}
\DeclareMathOperator*{\argmin}{arg\,min}

\title{Jealousy-freeness and other common properties \\ in Fair Division of Mixed Manna}

\author{
Martin Aleksandrov
\affiliations
TU Berlin, Germany
\emails
martin.aleksandrov@tu-berin.de
}

\begin{document}

\maketitle

\begin{abstract}
We consider a fair division setting where indivisible items are allocated to agents. Each agent in the setting has strictly negative, zero or strictly positive utility for each item. We, thus, make a distinction between items that are good for some agents and bad for other agents (i.e.\ mixed), good for everyone (i.e.\ goods) or bad for everyone (i.e.\ bads). For this model, we study axiomatic concepts of allocations such as \emph{jealousy-freeness up to one item}, \emph{envy-freeness up to one item} and \emph{Pareto-optimality}. We obtain many new possibility and impossibility results in regard to combinations of these properties. We also investigate new computational tasks related to such combinations. Thus, we advance the state-of-the-art in fair division of mixed manna.
\end{abstract}

\section{Introduction}\label{sec:intro}

Consider a Foodbank problem where donated food is given out to people in need. Such shelters exist in many countries around the world. Although some people would gladly accept any ``free-of-charge'' food, other people might find some of it undesirable. Thus, people may like some food items and dislike others. Also, consider a paper assignment problem where scientific papers are matched to reviewers. Although reviewers tend to bid for papers from their own field of expertise, they might as well have to bid for papers from other areas. Thus, they receive papers they like and also often papers they dislike. Both of these problems can be modeled mathematically as resource allocations of indivisible items to agents. 

Resource allocation of indivisible items lies on the intersection of fields such as social choice theory, computer science and algorithmic economics. Though a large body of work is devoted to the case when the items are goods (e.g.\ \cite{brams1996,moulin2003,steinhaus1948,young1995}), there is a rapidly growing interest in the case of mixed manna (e.g.\ \cite{aleksandrov2019greedy,aziz2019popropone,caragiannis2012,sandomirskiy2019minimal}). In a mixed manna, each item can be classified as \emph{mixed} (i.e.\ some agents strictly like it and other agents strictly dislike it), \emph{good} (i.e.\ all agents weakly like it and some agents strictly like it) or \emph{bad} (i.e.\ all agents weakly dislike it and some agents strictly dislike it). 

An allocation of the manna gives to each agent some different bundle of items. A common task in resource allocation is to compute an allocation that minimizes the inequalities between the agents' utilities for their bundles. A central axiomatic property that encodes this objective is equitability. An allocation is \emph{equitable} if all agents derive exactly the same utility from their bundles (i.e.\ perfect equitability). Unfortunately, such allocations might not exist even with two agents and one item. In response, we might consider approximate versions of equitability. For example, we might require that pairwise perfect equitability is restored whenever one item is moved across the agents' bundles. However, it might also not be possible to achieve this version simply because it requires that the agents' altered utilities become perfectly equal.

We receive an inspiration from the work of Gourv{\`{e}}s et al.\ \shortcite{gourves2014} in order to relax the ``perfect equitability" requirement. They proposed an axiomatic property such as jealousy-freeness which does precisely this. An agent is \emph{jealous} of another agent in a given allocation if the utility of the former agent for their own bundle is strictly lower than the utility of the latter agent for their own bundle. Otherwise, the former agent is \emph{jealousy-free} of the latter agent. For any pair of agents, it is always the case that one of them is jealousy-free of the other one. However, a given allocation is \emph{jealousy-free} if no agent is jealous of any other agent. Thus, such an allocation is also equitable. On the plus side, this means that there is \emph{zero} inequality in it. On the minus side, it also means that such an allocation might not exist.

\begin{table*}
\centering
\resizebox{0.65\textwidth}{2.55cm}{
\begin{tabular}{|c|c|c|C|C|C|C|}
\hline

{\bf properties} & {\bf agents} & {\bf items} & \multicolumn{4}{c|}{\bf manna with mixed items} \\ \hline

PO & $\geq 2$ & $\geq 1$ & \multicolumn{4}{c|}{$\checkmark$, leximin (Rem~\ref{rem:one})} \\ \hline

JF1 & $\geq 2$ & $\geq 3$ & \multicolumn{4}{c|}{$\times$ (Prop~\ref{pro:impmixed}$^{\star}$), $\NP$-hard (Thm~\ref{thm:hardmixed}$^{\star}$)} \\ \hline

EFX+PO & $2$ & $\geq 1$ & \multicolumn{4}{c|}{$\checkmark$, leximin (Thm~\ref{thm:efonetwo}$^{\star}$)}  \\ \hline

& & & \multicolumn{4}{c|}{\bf manna without mixed items}  \\ \cline{4-7}

&  & & \multicolumn{2}{c}{\bf goods and bads} & \multicolumn{2}{c|}{\bf pure goods and bads} \\ \hline

JFX$_0$; EFX$_0$ & $\geq 2$ & $\geq 3$ & \multicolumn{2}{c}{} & \multicolumn{2}{c|}{$\times$; $\times$ (Prop~\ref{pro:impefxjfxzero}$^{\star}$), open} \\ \hline

JF1$_0$ & $\geq 2$ & $\geq 1$ & \multicolumn{4}{c|}{$\checkmark$, Algorithm~\ref{alg:jfone} (Thm~\ref{thm:jfonealg})} \\ \hline

JFX & $\geq 2$ & $\geq 1$ & \multicolumn{4}{c|}{$\checkmark$, leximin$++$ (Thm~\ref{thm:jfxlex})} \\ \hline

JF1+PO & $\geq 2$ & $\geq 4$ & \multicolumn{2}{c|}{$\times$ (Prop~\ref{pro:impall}$^{\star}$), $\NP$-hard (Thm~\ref{thm:hardgoods}$^{\star}$)} & \multicolumn{2}{c|}{} \\ \cline{1-3}

JFX+PO & $\geq 2$ & $\geq 1$ & \multicolumn{2}{c|}{} & \multicolumn{2}{c|}{$\checkmark$, leximin (Cor~\ref{cor:jfxone})} \\ \cline{1-3} \cline{6-7}

JFX+EFX+PO & $2$ & $\geq 1$ & \multicolumn{2}{c|}{} & \multicolumn{2}{c|}{$\checkmark$, leximin (Cor~\ref{cor:jfxtwo}$^{\star}$)} \\ \cline{1-5}

JFX+EFX & $2$ & $\geq 1$ & \multicolumn{2}{c|}{$\checkmark$, leximin$++$ (Thm~\ref{thm:efxjfx}$^{\star}$)} &\multicolumn{2}{c|}{} \\ \hline

JF1+EF1 & $\geq 2$ & $\geq 2$ & \multicolumn{2}{c}{} & \multicolumn{2}{c|}{$\times$ (Prop~\ref{pro:impnorm}), open} \\ \hline

JF1+EF1 & $\geq 2$ & $\geq 4$ & \multicolumn{2}{c}{} & \multicolumn{2}{c|}{$\times$ (Prop~\ref{pro:impjfxefx}$^{\star\star}$), open} \\ \hline

JF1+EF1 & $\geq 3$ & $\geq 3$ & \multicolumn{2}{c}{} & \multicolumn{2}{c|}{$\times$ (Prop~\ref{pro:imphouse}$^{\star}$), open} \\ \hline

\end{tabular}
}
\captionsetup{justification=centering}
\caption{Results for $n\geq 2$ agents: $\checkmark$-possible, $\times$-not possible, $\star$-normalised additive utilities, $\star\star$-normalised general utilities.}
\label{tab:results}
\end{table*}

\begin{figure*}
\centering
\subfigure[mixed manna]{\includegraphics[scale=0.2]{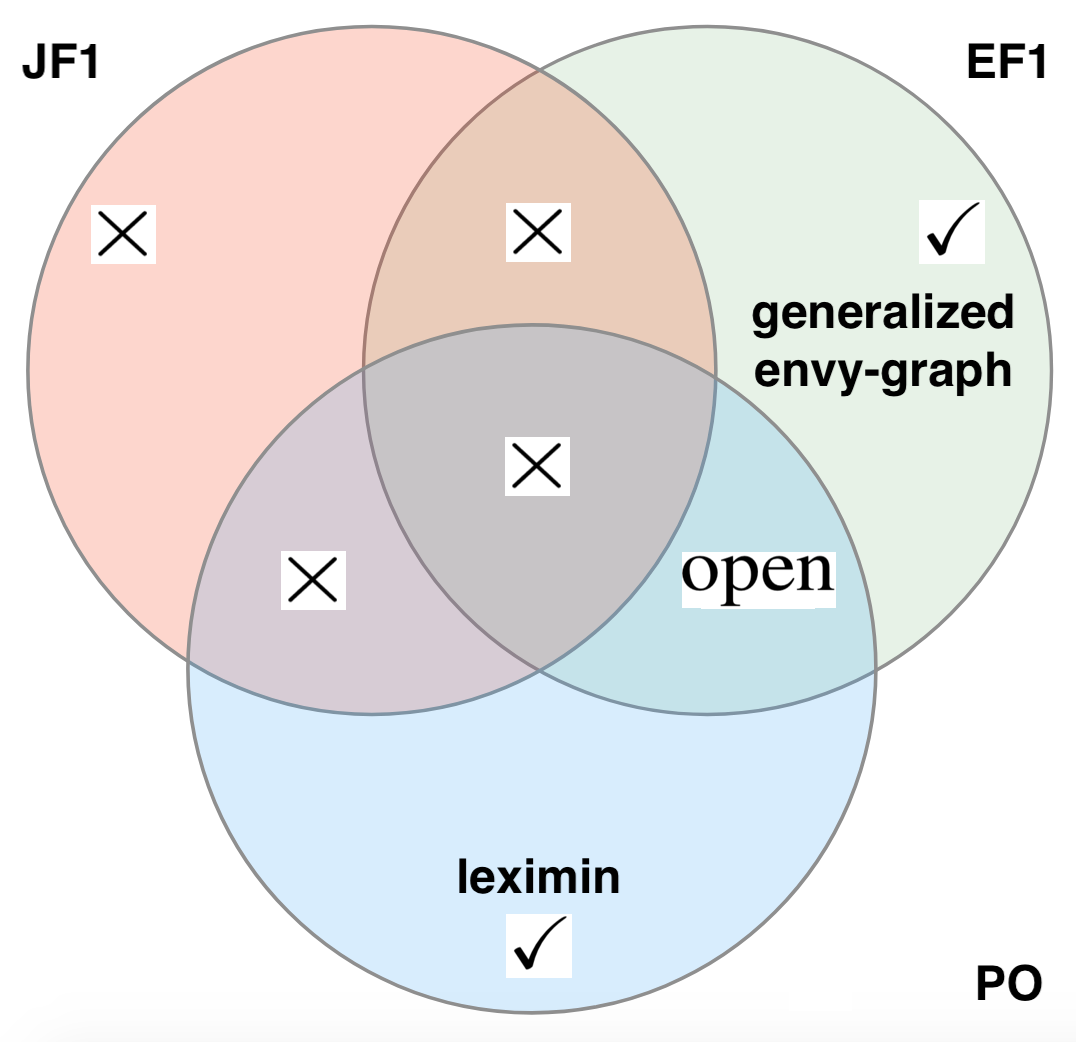}\label{fig:mixed}}
\centering
\subfigure[goods \& bads]{\includegraphics[scale=0.2]{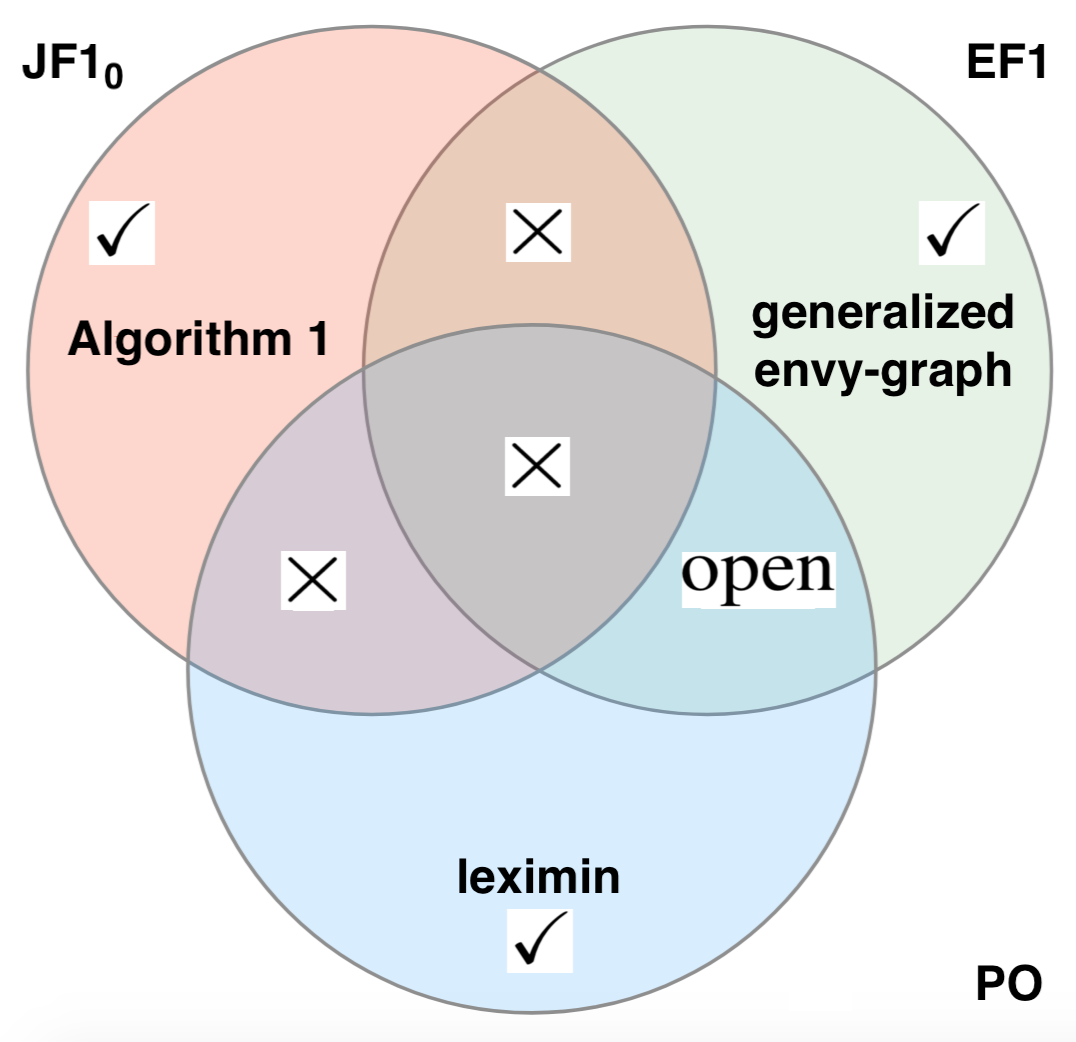}\label{fig:gbfirst}}
\centering
\subfigure[goods \& bads]{\includegraphics[scale=0.2]{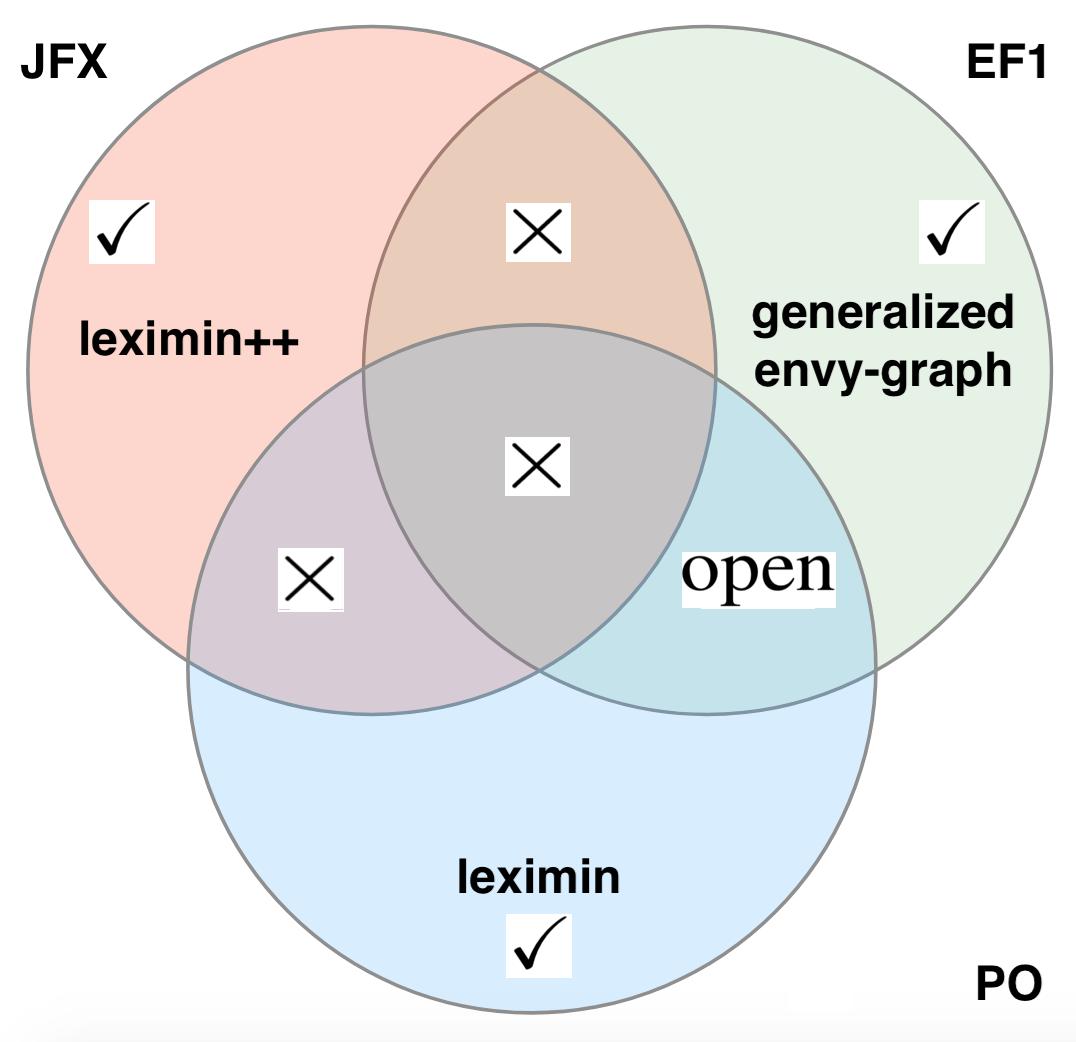}\label{fig:gbsecond}}
\centering
\subfigure[pure goods \& bads]{\includegraphics[scale=0.2]{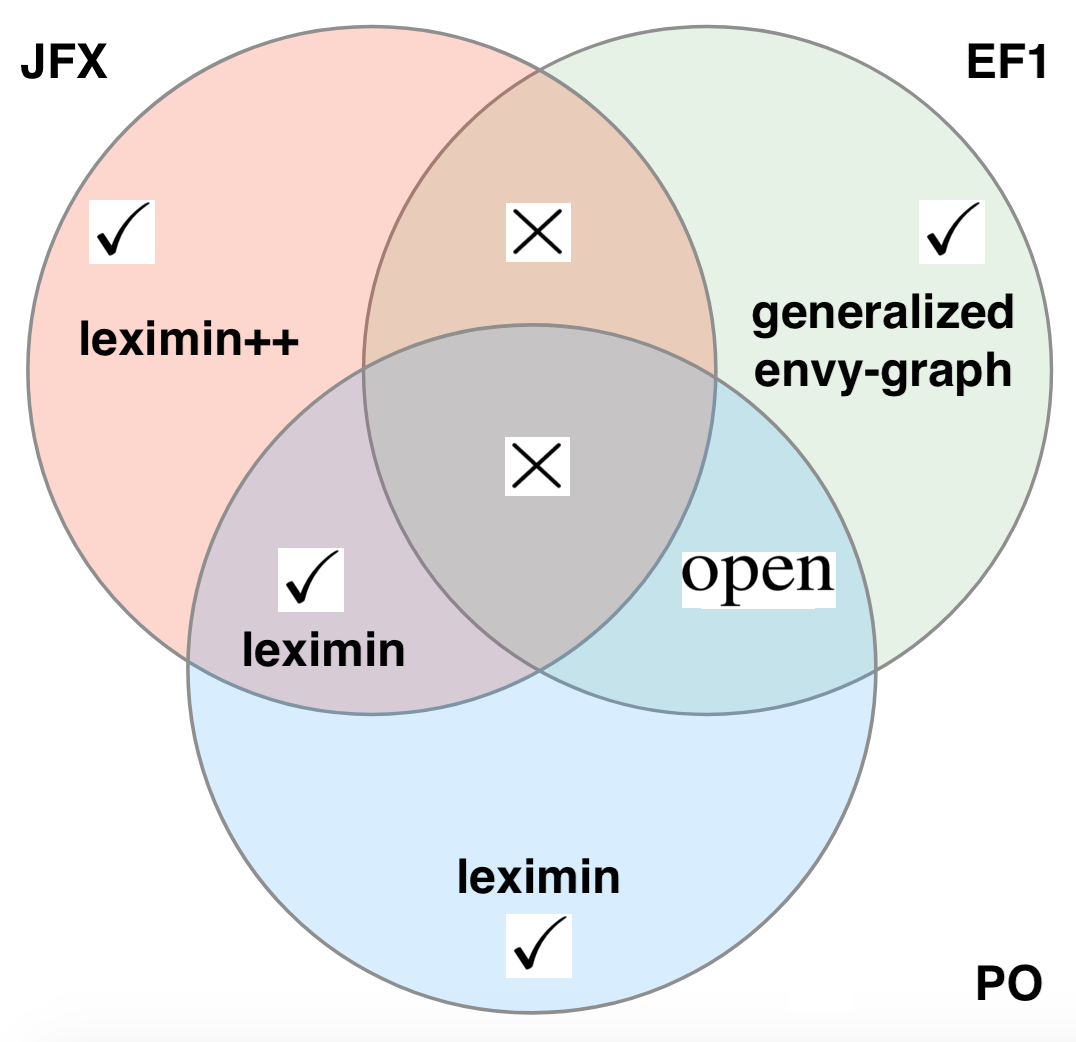}\label{fig:pgb}}
\captionsetup{justification=centering}
\vspace{-0.15cm}
\caption{Combinations for $n\geq 2$ agents: the generalized envy-free algorithm from \protect\cite{aziz2019gc}, \\ leximin from \protect\cite{dubins1961}, leximin++ from \protect\cite{plaut2018}, Algorithm~\ref{alg:jfone} from our work.}  
\label{fig:results}
\end{figure*}

In this paper, we analyse four new relaxations of jealousy-freeness (i.e.\ JF1, JF1$_0$, JFX and JFX$_0$) for allocations of the manna. They insist on achieving jealousy-freeness by decreasing the utilities of jealousy-free agents in ``up to one item'' fashion. For example, an allocation is JF1 if each agent is jealousy-free of any other agent or, otherwise, an agent who is jealous becomes jealousy-free of another agent, after some non-zero valued bad from the jealous agent's bundle is added to the other agent's bundle \emph{or} some non-zero valued good is removed from the other agent's bundle. JFX strengthens these requirements to each bad in the jealous agent's bundle and each good in the other agent's bundle. Furthermore, JF1$_0$ and JFX$_0$ relax the ``non-zero valued'' requirements imposed by JF1 and JFX, respectively. We will shortly observe the following relations between these properties.

\begin{center}
JFX$_0$\hspace{0.25cm}$\Rightarrow$\hspace{0.25cm}JFX\hspace{0.25cm}$\Rightarrow$\hspace{0.25cm}JF1$_0$\hspace{0.25cm}$\Rightarrow$\hspace{0.25cm}JF1
\end{center} 

We also investigate how these properties interact with common efficiency and fairness criteria such as Pareto-optimality (PO) and envy-freeness up to one item (EF1, EFX and EFX$_0$). PO ensures that we cannot re-distribute items among agents' bundles so that we make every agent weakly happier and some agent strictly happier. EF1 and EFX for our model are from \cite{aziz2019gc}. For example, EF1 requires that an agent's envy for another agent's bundle is eliminated by removing some item from these agents' bundles. EFX strengthens EF1 to any non-zero valued item in these bundles, increasing the envy agent's utility or decreasing the other agent's utility. EFX$_0$ extends envy-freeness up to any (possibly zero valued) good from \cite{plaut2018} to any (possibly zero valued) bad. These fairness properties obey the following well-known pattern.

\begin{center}
EFX$_0$\hspace{0.25cm}$\Rightarrow$\hspace{0.25cm}EFX\hspace{0.25cm}$\Rightarrow$\hspace{0.25cm}EF1
\end{center}

\section{Our contributions}\label{sec:ourresults}

\begin{figure}[h]
\centering
\subfigure[mixed manna]{\includegraphics[scale=0.2]{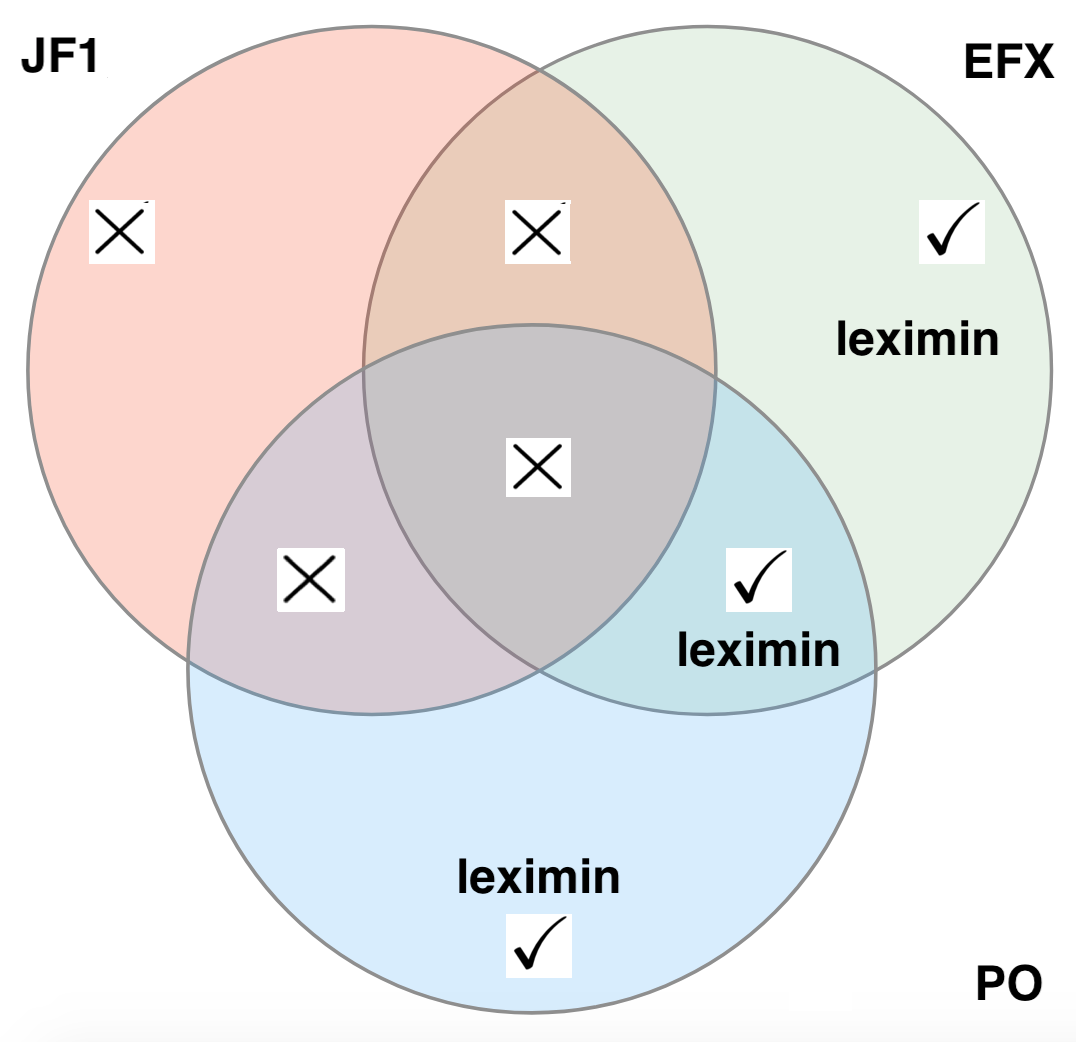}\label{fig:mixedtwo}}
\centering
\subfigure[goods \& bads]{\includegraphics[scale=0.2]{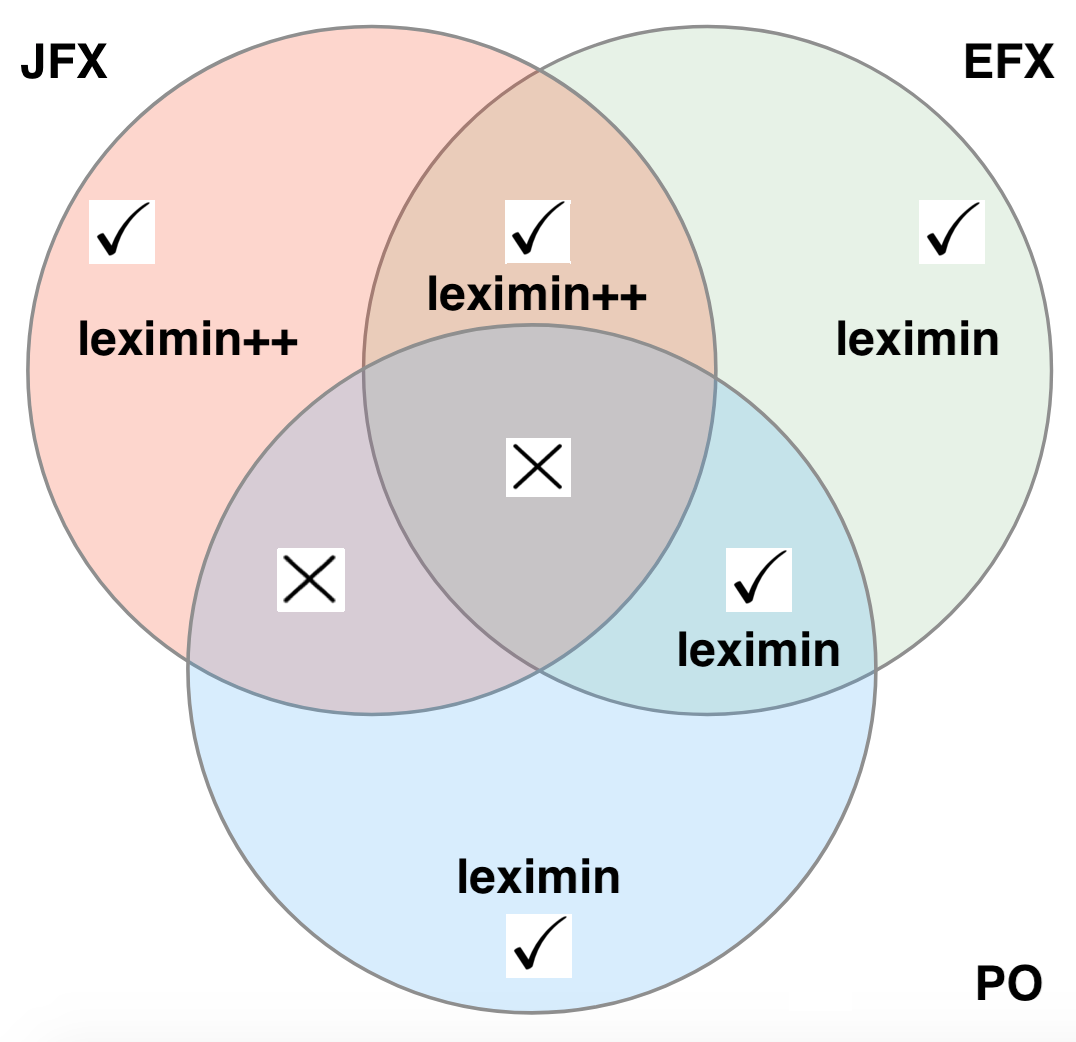}\label{fig:gbtwo}}\\
\subfigure[pure goods \& bads]{\includegraphics[scale=0.2]{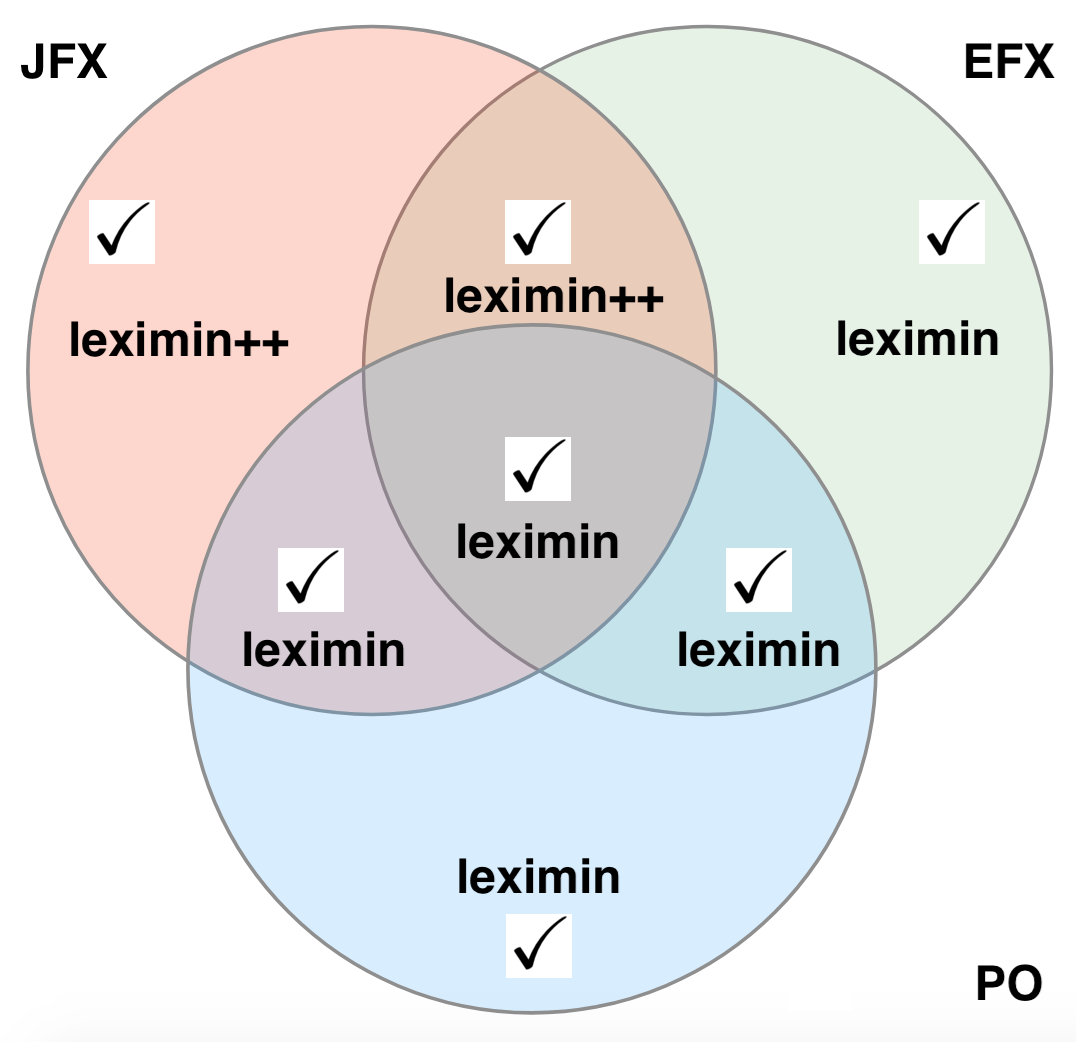}\label{fig:pgbtwo}}
\captionsetup{justification=centering}
\vspace{-0.15cm}
\caption{Combinations for $2$ agents and norm. additive utilities.}  
\label{fig:resultstwo}
\end{figure}

We highlight in this section our results. Although we emphasise in our work on possibility and impossibility results, we report some computational results as well. 
Table~\ref{tab:results} contains all results. Figures~\ref{fig:results} and~\ref{fig:resultstwo} depict most of them and some existing results. Some possibility results rely on exponential-time solutions such as \emph{leximin} and \emph{leximin}$++$. Others rely on polynomial-time solutions such as Algorithm~\ref{alg:jfone}. In particular, our findings give answers to the following question regarding combinations of JF1, EF1 and PO.

\begin{quotation}
\em Question: When does it exist an allocation that satisfies a given combination of JF1, EF1 and PO?
\end{quotation}

We first report our results for the new properties JF1, JF1$_0$, JFX and JFX$_0$. 

\begin{itemize}
\item We start by proving that JF1 allocations might not exist in problems with mixed items. (Proposition~\ref{pro:impmixed}). The related decision problem is $\NP$-hard (Theorem~\ref{thm:hardmixed}). 

\item By comparison, JFX$_0$ allocations might not exist in problems without mixed items (Proposition~\ref{pro:impefxjfxzero}). 

\item Further, we show that each leximin$++$ allocation in such problems is JFX (Theorem~\ref{thm:jfxlex}).

\item We also show that Algorithm~\ref{alg:jfone} returns an JF1$_0$ allocation in such problems (Theorem~\ref{thm:jfonealg}).
\end{itemize}

We further summarize our results for combinations, including PO but excluding EF1, EFX and EFX$_0$.

\begin{itemize}
\item The leximin solution satisfies PO (Remark~\ref{rem:one}).

\item JF1 and PO might be incompatible in problems with goods and bads (Proposition~\ref{pro:impall}). The related decision problem is $\NP$-hard (Theorem~\ref{thm:hardgoods}). 

\item Nevertheless, each leximin allocation of pure goods (i.e.\ all agents strictly like them) and bads is JFX and PO (Corollary~\ref{cor:jfxone}).
\end{itemize}

We lastly list our results for other combinations, including also EF1, EFX and EFX$_0$. Some of these add solely to the literature of EF1, EFX, EFX$_0$ and PO. 

\begin{itemize}
\item For the case of \num{2} agents, EFX$_0$ allocations might not exist in problems without mixed items (Proposition~\ref{pro:impefxjfxzero}). 

\item On the contrary, each leximin allocation is EFX and PO with normalised additive utilities for any manna (Theorem~\ref{thm:efonetwo}), and JFX for pure goods and bads (Corollary~\ref{cor:jfxtwo}).  

\item We also show that each leximin$++$ allocation is JFX and EFX with normalised additive utilities for goods and bads (Theorem~\ref{thm:efxjfx}).

\item We additionally prove that JF1 and EF1 are incompatible in two contexts with \num{2} agents (Propositions~\ref{pro:impnorm}-\ref{pro:impjfxefx}) and one context with \num{3} agents (Proposition~\ref{pro:imphouse}).
\end{itemize}

Although leximin and leximin$++$ are exponential-time solutions, they often come with nice axiomatic guarantees. For this reason, we feel that their computability would not be problematic in settings with few items \cite{bliem2016}.

\section{Related work}\label{sec:rel}

Jealousy-freeness up to one item relates to \emph{near} jealousy-freeness for matroids from \cite{gourves2014}. In fact, JFX$_0$ and JFX coincide with this notion in their setting. Hence, their near jealousy-freeness algorithm returns JFX$_0$ allocations in problems with additive utilities for pure goods. By comparison, we prove that such allocations stop to exist as soon as we add bads to the problem. Other works of matroids are \cite{gourves2013} and \cite{gourves2013wine}. They also consider only positive-valued utility functions whereas we consider real-valued utility functions.

Jealousy-freeness up to one item appears to be an ideal complement to minimizing inequality. To elaborate further on this, let us consider the popular Gini index \cite{gini1912}. Aleksandrov and Walsh \cite{aleksandrov2019epia} showed that each allocation that minimizes the Gini index in some problems with pure goods might not share any PO guarantees. In contrast, we prove that JFX and PO allocations always exist in problems with pure goods and bads. Other works of inequalities are \cite{endriss2013} and \cite{schneckenburger2017}. However, they consider inequality measures for allocations of goods whereas we consider inequality properties for allocations of goods, bads and mixed items.

JF1$_0$ and JFX relate to existing approximations of equitability. Freeman et al.\ \shortcite{freeman2019ijcai} proposed two such notions in the context of goods (i.e.\ EQ1 and EQX). They proved that the leximin solution is EQX and PO in problems with pure goods. They also studied EQ1 and EQX in the context of bads \cite{freeman2020equitable}. However, they discovered that this solution is no longer EQ1 and no other allocation is EQX and PO. In response, they proposed two notions of equitability up to one duplicated item (i.e.\ DEQ1 and DEQX). Each of them is compatible with PO. Notably, EQ1, EQX for goods and DEQ1, DEQX for bads require like our properties that jealousy-freeness is restored by diminishing the jealousy-free agents' utilities. As a result, JF1$_0$ and JFX degenerate respectively to EQ1, EQX for goods and DEQ1, DEQX for bads.

Jealousy-freeness up to one item does not relate much to EF1. For example, Aziz et al.\ \shortcite{aziz2019gc} gave the generalized envy-graph algorithm for computing EF1 allocations. At the same time, we show that JF1 allocations might not exist in problems with mixed items and prove that the related computational problem is $\NP$-hard. For problems without mixed items, we give a greedy algorithm for computing JF1$_0$ allocations. Nevertheless, even with pure goods in the problem, we prove that JF1 and EF1 cannot be achieved together. As a result, other existing EF1 approaches (e.g.\ the MNW solution from \cite{caragiannis2016}) may fail JF1 and also new JF1 approaches (e.g.\ our greedy algorithm) may fail EF1.

Jealousy-freeness up to one item relates even less to EFX and EFX$_0$. Caragiannis et al.\ \shortcite{caragiannis2016} proposed EFX for goods. Although EF1 allocations exist, it remained an open question in the last years whether EFX allocations exist in general. We close the case of 2 agents for our model with normalised additive utilities. For example, the leximin allocation is EFX and PO in such problems whereas the leximin$++$ allocation is EFX and JFX in such problems without mixed items. Plaut and Roughgarden \shortcite{plaut2018} considered EFX$_0$ even though Kyropoulou et al.\ \shortcite{kyropoulou2019} coined the name. EFX$_0$ allocations exist in problems with \num{2} agents and goods. This may not be true in our setting even with just \num{2} agents.

Seemingly, some of our results rely on the leximin$++$ solution \cite{plaut2018}. This solution has received less attention than the leximin solution (e.g.\ \cite{rawls1971,sen1976,sen1977}). The leximin solution is PO. However, we give problems with goods where it may fail JF1 whereas the leximin$++$ solution is JFX even with bads but it may not be PO. Computing the leximin and leximin$++$ solutions relates to computing max-min fair allocations which is $\NP$-hard \cite{bezakova2005ind,dobzinski2013}. Other related works are listed in \cite{freeman2019ijcai,freeman2020equitable} and \cite{aziz2019gc}. However, our results do not follow from existing results.

\section{Formal preliminaries}\label{sec:pre}

In this section, we define more formally the model of fair division of mixed manna, the aforementioned axiomatic properties for allocations in this model as well as the leximin and leximin$++$ solutions.

\subsection{Model}\label{subsec:model}

We consider a set $[n]=\lbrace 1,\ldots, n\rbrace$ of $n\in\mathbb{N}_{\geq 2}$ agents and a set $[m]=\lbrace 1,\ldots,m\rbrace$ of $m\in\mathbb{N}_{\geq 1}$ indivisible items. We let each $a\in [n]$ have some \emph{utility} function $u_a:2^{[m]}\rightarrow\mathbb{R}$ Thus, they assign some bundle utility $u_a(M)$ to each bundle $M\subseteq [m]$. We write $u_a(o)$ for $u_a(\lbrace o\rbrace)$. We say that $u_a(M)$ is \emph{additive} iff $u_a(M)=\sum_{o\in M} u_a(o)$. Otherwise, we say that it is \emph{general}. Also, we sometimes consider \emph{normalised} utilities. In this case, we suppose that $u_a(\emptyset)=0$ and $u_a([m])=c$ holds for each $a\in [n]$ and some $c\in\mathbb{R}$.

With additive utilities, the set of items $[m]$ can be partitioned into \emph{mixed items}, \emph{goods} and \emph{bads}. Respectively, we write $[m]^{\pm}=\lbrace o\in [m]|\exists a\in [n]: u_a(o)>0,\exists b\in [n]:u_b(o)<0\rbrace$, $[m]^+=\lbrace o\in [m]|\forall a\in [n]: u_a(o)\geq 0,\exists b\in [n]:u_b(o)>0\rbrace$ and $[m]^-=\lbrace o\in [m]|\forall a\in [n]:u_a(o)\leq 0,\exists b\in [n]:u_b(o)<0\rbrace$ for the sets of these items. We refer to an item $o$ from $[m]^+$ as a \emph{pure good} if $\forall a\in [n]: u_a(o)>0$. Also, we refer to an item $o$ from $[m]^-$ as a \emph{pure bad} if $\forall a\in [n]: u_a(o)<0$. 

With additive utilities, the type of item (i.e.\ mixed, good or bad) does not depend on the distribution of bundles to agents but only on the combination of the agents' cardinal utilities for the item. With general utilities, we cannot do this partitioning because now whether a given item is mixed, good or bad depends on the distribution of bundles to agents and the combination of the agents' marginal utilities for the item with respect to their bundles. We next illustrate this.

\begin{myexample}\label{exp:gen}
Suppose there are \num{2} agents and \num{4} items, say $a$, $b$, $c$ and $d$. Further, consider the following utilities for some of the bundles of these items.

\begin{center}
\begin{tabular}{|c|c|c|c|} \hline
   bundle & utility & bundle & utility \\ \hline
   $\lbrace b\rbrace$ & $1$ & $\lbrace c\rbrace$ & $3$ \\
   $\lbrace a,b\rbrace$ & $2$ & $\lbrace a,c\rbrace$ & $2$ \\
   $\lbrace b,d\rbrace$ & $2$ & $\lbrace c,d\rbrace$ & $2$ \\
   $\lbrace a,b,d\rbrace$ & $1.5$ & $\lbrace a,c,d\rbrace$ & $4$ \\ \hline
\end{tabular}
\end{center}

We can observe that an item can be good in one allocation and bad in another one. To see this, pick $A=(\lbrace a,b\rbrace,\lbrace c,d\rbrace)$ and $B=(\lbrace a,c\rbrace,\lbrace b,d\rbrace)$ and focus on item $a$. This item is pure good in $A$ because $u_1(A_1)-u_1(A_1\setminus\lbrace a\rbrace)=1>0$ and $u_2(A_2\cup\lbrace a\rbrace)-u_2(A_2)=2>0$ hold. However, item $a$ is pure bad in $B$ because $u_1(B_1)-u_1(B_1\setminus\lbrace a\rbrace)=-1<0$ and $u_2(B_2\cup\lbrace a\rbrace)-u_2(B_2)=-0.5<0$ hold.\myqed
\end{myexample}

In this case, let us consider agent $a\in [n]$, item $o\in [m]$ and bundle $M\subseteq [m]\setminus\lbrace o\rbrace$. We say that $o$ is \emph{good} for $a$ with respect to $M$ if $u_a(M\cup\lbrace o\rbrace)\geq u_a(M)$. We refer to $o$ as \emph{pure good} whenever $u_a(M\cup\lbrace o\rbrace)>u_a(M)$. Similarly, we say that $o$ is \emph{bad} for $a$ with respect to $M$ if $u_a(M\cup\lbrace o\rbrace)\leq u_a(M)$. We refer to $o$ as \emph{pure bad} whenever $u_a(M\cup\lbrace o\rbrace)<u_a(M)$. Further, let us consider another agent $b\in [n]$ and another bundle $N\subseteq [m]\setminus(M\cup\lbrace o\rbrace)$. Thus, we say that $o$ \emph{mixed} if $u_a(M\cup\lbrace o\rbrace)>u_a(M)$ and $u_b(N\cup\lbrace o\rbrace)<u_b(N)$.


We pay a special attention in our work to three types of problems. In a problem \emph{with} mixed items, there is an allocation, an item and two agents such that one of the agents' marginal utilities for the item in the allocation is strictly positive and the other one is strictly negative. In a problem \emph{without} mixed items, all agents reach a consensus on whether a given item is good or bad in a given allocation. In a problem with \emph{pure goods} and \emph{bads}, the agents' marginal utilities for a given item in a given allocation are either all strictly positive or all weakly negative. 

\subsection{Axiomatic properties}\label{subsec:axioms}

An \emph{(complete) allocation} $A=(A_1,\ldots,A_n)$ is such that (1) $A_a$ is the set of items allocated to agent $a\in [n]$, (2) $\cup_{a\in [n]} A_a=[m]$ and (3) $A_a\cap A_b=\emptyset$ for each $a,b\in[n]$ with $a\neq b$. We consider several properties for allocations.

\paragraph{Jealousy-freeness up to one item}\label{par:jfone} Let us consider an allocation and a pair of agents, say 1 and 2. One of them is jealousy-free of the other one, say 2. Our approximations of jealousy-freeness rely on the idea of decreasing the utility of the agent who is jealousy-free, i.e.\ 2's utility.

Thus, agent 1 is JF1 of agent 2 whenever 1's utility is at least as much as 2's utility, after taking a non-zero valued bad from 1's bundle and \emph{adding} it to 2's bundle or \emph{removing} a non-zero valued good from 2's bundle.

\begin{mydefinition} $(${\em JF1}$)$
An allocation $A$ is \emph{jealousy-free up to some non-zero valued item} if, $\forall a,b\in [n]$, $u_a(A_a)\geq u_b(A_b)$, $\exists o\in A_a$ s.t. $u_a(A_a)<u_a(A_a\setminus\lbrace o\rbrace)$: $u_a(A_a)\geq u_b(A_b\cup\lbrace o\rbrace)$ or $\exists o\in A_b$ s.t. $u_b(A_b)>u_b(A_b\setminus\lbrace o\rbrace)$: $u_a(A_a)\geq u_b(A_b\setminus\lbrace o\rbrace)$.
\end{mydefinition} 

Agent 1 is JFX of agent 2 whenever the above requirements hold for any bad in 1's bundle, strictly increasing 1's utility, and any good in 2's bundle, strictly decreasing 2's utility. 

\begin{mydefinition}$(${\em JFX}$)$
An allocation $A$ is \emph{jealousy-free up to any non-zero valued item} if, $\forall a, b\in [n]$, (1) $\forall o\in A_a$ s.t. $u_a(A_a)<u_a(A_a\setminus\lbrace o\rbrace)$: $u_a(A_a)\geq u_b(A_b\cup\lbrace o\rbrace)$ and (2) $\forall o\in A_b$ s.t. $u_b(A_b)>u_b(A_b\setminus\lbrace o\rbrace)$: $u_a(A_a)\geq u_b(A_b\setminus\lbrace o\rbrace)$.
\end{mydefinition} 

JF1 is a strictly weaker concept than JFX. Indeed, there are problems where a JF1 allocation might violate JFX. Interestingly, this can be observed even in problems where agents have the same utility for each item.

\begin{myexample}\label{exp:rel}
Let us consider a problem with \num{2} agents and \num{3} pure bads, subject to the utilities in the below matrix. 

\begin{center}
\begin{tabular}{|c|c|c|c|} \hline
   & a & b & c \\ \hline
   agent 1 & $-1$ & $-2$ & $-3$ \\
  agent 2 & $-1$ & $-2$ & $-3$ \\ \hline
\end{tabular}
\end{center}

The allocation $A=(\lbrace a,c\rbrace,\lbrace b\rbrace)$ is such that $u_1(A_1)=-4<-3=u_2(A_2\cup\lbrace a\rbrace)$ and $u_1(A_1)=-4>-5=u_2(A_2\cup\lbrace c\rbrace)$ hold. Hence, $A$ is JF1 but not JFX.\myqed
\end{myexample}

By comparison, an allocation that satisfies JFX is clearly JF1. This follows directly by these concepts' definitions. 

Freeman et al.\ \shortcite{freeman2019ijcai} considered a stronger variant of JF1 for problems with goods, not-imposing the non-zero marginal requirements. We generalize this concept to our setting in a similar fashion. 

\begin{mydefinition} $(${\em JF1$_0$}$)$
An allocation $A$ is \emph{jealousy-free up to some item} if, $\forall a,b\in [n]$, $u_a(A_a)\geq u_b(A_b)$, $\exists o\in A_a$: $u_a(A_a)\geq u_b(A_b\cup\lbrace o\rbrace)$ or $\exists o\in A_b$: $u_a(A_a)\geq u_b(A_b\setminus\lbrace o\rbrace)$.
\end{mydefinition} 

Freeman et al.\ \shortcite{freeman2020equitable} also defined similarly a stronger notion than JFX for problems with bads. We generalize this concept to our setting by relaxing the non-zero marginal requirements and refer to it as JFX$_0$.

\begin{mydefinition}$(${\em JFX$_0$}$)$
An allocation $A$ is \emph{jealousy-free up to any item} if, $\forall a, b\in [n]$, (1) $u_a(A_a)\geq u_b(A_b\cup\lbrace o\rbrace)$ for each $o\in A_a$ s.t. $u_a(A_a)\leq u_a(A_a\setminus\lbrace o\rbrace)$ and (2) $u_a(A_a)\geq u_b(A_b\setminus\lbrace o\rbrace)$ for each $o\in A_b$ s.t. $u_b(A_b)\geq u_b(A_b\setminus\lbrace o\rbrace)$.
\end{mydefinition} 

A JFX$_0$ allocation is also JFX. Moreover, JFX is stronger than JF1$_0$ and JF1$_0$ is stronger than JF1. These relations follow directly by the definitions of these concepts.

\paragraph{Envy-freeness up to one item}\label{par:ef}

Envy-freeness up to one item requires that an agent's envy for another's bundle is eliminated by removing an item from the bundles of these agents. Two notions for our model that are based on this idea are EF1 and EFX \cite{aziz2019gc}.

\begin{mydefinition} $(${\em EF1}$)$
An allocation $A$ is \emph{envy-free up to some item} if, $\forall a,b\in [n]$, $u_a(A_a)\geq u_a(A_b)$ or $\exists o\in A_a\cup A_b$ s.t. $u_a(A_a\setminus\lbrace o\rbrace)\geq u_a(A_b\setminus\lbrace o\rbrace)$.
\end{mydefinition} 

\begin{mydefinition}$(${\em EFX}$)$
An allocation $A$ is \emph{envy-free up to any non-zero valued item} if, $\forall a, b\in [n]$, (1) $\forall o\in A_a$ s.t. $u_a(A_a)$ $<$ $u_a(A_a\setminus\lbrace o\rbrace)$: $u_a(A_a\setminus\lbrace o\rbrace)\geq u_a(A_b)$ and (2) $\forall o\in A_b$ s.t. $u_a(A_b)>u_a(A_b\setminus\lbrace o\rbrace)$: $u_a(A_a)\geq u_a(A_b\setminus\lbrace o\rbrace)$.
\end{mydefinition} 

Plaut and Roughgarden \shortcite{plaut2018} considered a variant of EFX for goods where, for any given pair of agents, the removed item may be valued with zero utility by the envy agent. Kyropoulou et al.\ \shortcite{kyropoulou2019} referred to this one as EFX$_0$. We adapt this property to our model by relaxing the non-zero marginal requirements in the definition of EFX. 

\begin{mydefinition}$(${\em EFX$_0$}$)$
An allocation $A$ is \emph{envy-free up to any item} if, $\forall a, b\in [n]$, (1) $u_a(A_a\setminus\lbrace o\rbrace)\geq u_a(A_b)$ for each $o\in A_a$ s.t. $u_a(A_a)$ $\leq$ $u_a(A_a\setminus\lbrace o\rbrace)$ and (2) $u_a(A_a)\geq u_a(A_b\setminus\lbrace o\rbrace)$ for each $o\in A_b$ s.t. $u_a(A_b)\geq u_a(A_b\setminus\lbrace o\rbrace)$.
\end{mydefinition} 

An allocation that is EFX$_0$ further satisfies EFX. Also, EFX is stronger than EF1. It is well-known that the opposite relations might not hold. 

\paragraph{Pareto-optimality}\label{par:po}

Vilfredo Pareto had proposed its optimality a long time ago in his seminal work \cite{pareto1896}. We next define it formally for allocations in our model. 

\begin{mydefinition}$(${\em PO}$)$
An allocation $A$ is \emph{Pareto-optimal} if there is no allocation $B$ that \emph{Pareto-improves} $A$, i.e.\ $\forall a\in [n]$: $u_a(B_a)\geq u_a(A_a)$ and $\exists b\in [n]$: $u_b(B_b)> u_b(A_b)$.
\end{mydefinition} 

\subsection{Leximin and leximin$++$}\label{subsec:solutions}

Plaut and Roughgarden \shortcite{plaut2018} implemented one operator for comparing allocations: $\succ$. This operator induces a total order between allocations. Thus, an \emph{leximin} allcation is a maximal element under this order. Such an allocation maximizes the minimum utility of any agent, subject to which the second minimum utility is maximized, and so on. For this reason, each leximin allocation is trivially PO. 

\begin{myremark}\label{rem:one}
In fair division of mixed manna with general utilities, the leximin solution is PO.
\end{myremark}

Plaut and Roughgarden \shortcite{plaut2018} further proposed another total operator for comparing allocations: $\succ_{++}$ They refer to the maximal elements under it as \emph{leximin}$++$ allocations. Such an allocation maximizes the minimum utility, then maximizes the size of the bundle of an agent with minimum utility, before it maximizes the second minimum utility and the size of the second minimum utility bundle, and so on. Unfortunately, leximin$++$ allocations might generally not be PO.

Freeman et al.\ \shortcite{freeman2020equitable} noted that there might multiple leximin allocations in some problems. Perhaps, the most relavant to us is that this observation holds for leximin$++$ allocations as well (see Example~\ref{exp:rel}). 

\section{Common assumptions}\label{sec:ass}

An allocation in a problem with non-normalised utilities is envy-free up to one item or Pareto-optimal if and only if it is envy-free up to one item or Pareto-optimal in the corresponding problem with normalised utilities. This might not be true for a concept such as jealousy-freeness up to one item because normalisation reduces the agents' total utilities.

\begin{myexample}\label{exp:normone}
Let us consider the first problem with \num{4} goods and \num{2} agents. The second problem is its normalised version.

\begin{center}
\begin{tabular}{|c|c|c|c|c|} \hline
   & a & b & c & d \\ \hline
   & \multicolumn{4}{c|}{non-normalised} \\ \hline
  agent 1 & $1$ & $1$ & $1$ & $1$ \\
  agent 2 & $1$ & $0$ & $0$ & $0$ \\ 
  \hline
   & \multicolumn{4}{c|}{normalised} \\ \hline
     agent 1 & $\frac{1}{4}$ & $\frac{1}{4}$ & $\frac{1}{4}$ & $\frac{1}{4}$ \\
  agent 2 & $1$ & $0$ & $0$ & $0$ \\ \hline
\end{tabular}
\end{center}

The only JF1 and PO allocation with normalised utilities is $A=(\lbrace b,c,d\rbrace,\lbrace a\rbrace)$: agent 2 gets utility \num{1} and agent 1 gets utility $\frac{3}{4}$. However, $A$ falsifies JF1 with non-normalised utilities: agent 1 gets $1$ and agent 2 gets $3$. 
\myqed
\end{myexample}  

Additionally, an allocation that ignores agents with zero utilities for items is envy-free up to one item or Pareto-optimal wrt the ignored agents. Again, this may not be true for jealousy-freeness up to one item. In fact, the only way to achieve this concept in some problems might be to give items to such agents.

\begin{myexample}\label{exp:normtwo}
Let us consider the below problem with \num{4} goods and \num{2} agents, having 0/1 utilities for the items.

\begin{center}
\begin{tabular}{|c|c|c|c|c|} \hline
   & a & b & c & d \\ \hline
  agent 1 & $1$ & $1$ & $1$ & $1$ \\
  agent 2 & $0$ & $0$ & $0$ & $0$ \\ \hline
\end{tabular}
\end{center}

If we ignore agent 2, allocating all items to agent 1 is EF1 and PO but it is not even JF1 because agent 2 is not JF1 of agent 1. Otherwise, allocating one item to agent 1 and three items to agent 2 is JF1 but it is clearly neither EF1 nor PO. 
\myqed
\end{myexample}  

We conclude that common assumptions such as normalised utilities and ignoring agents with zero utilities might be too strong in our study. For this reason, we do \emph{not} make any of them throughout our work unless we explicitly mention it.

\section{JF1 with mixed items}\label{sec:jfonemix}

Consider again the Foodbank problem from the beginning of the paper. In this particular context, JF1 would somehow minimize the inequalities between the people's levels of satisfaction with the received food. Unfortunately, there are such settings where \emph{none} of the allocations is JF1 because some agent like items that another dislike. Thus, the utility levels of such agents in each allocation diverge from each other. 

\begin{myproposition}\label{pro:impmixed}
There are problems with \num{2} agents and normalised additive utilities for \num{2} mixed items and \num{1} bad, in which \emph{no} allocation is JF1.
\end{myproposition}  

\begin{myproof}
Let us consider a problem with \num{2} mixed items, \num{1} bad and \num{2} agents, having the following normalised utilities. 

\begin{center}
\begin{tabular}{|c|c|c|c|} \hline
   & a & b & c  \\ \hline
  agent 1 & $1$ & $1$ & $-4$ \\
  agent 2 & $-1$ & $-1$ & $0$  \\ \hline
\end{tabular}
\end{center}

Assume that a JF1 allocation exist in this problem. We let $A$ denote such an allocation. We derive a contradiction. 

\emph{Case 1}: Let $|\lbrace a,b\rbrace\cap A_1|\geq 2$ hold. If $c\in A_2$, then $u_1(A_1)=2$ whereas $u_2(A_2)=0$. Hence, $A$ cannot be JF1 because $u_2(A_2)<1\leq u_1(A_1\setminus\lbrace o\rbrace)$ for each $o\in \lbrace a,b\rbrace\cap A_1$. If $c\in A_1$, then $u_1(A_1)=-2$ whereas $u_2(A_2)=0$. Now, $A$ cannot be JF1 as well because of $u_2(c)=0$ and, therefore, $u_1(A_1)<0\leq u_2(A_2)=u_2(A_2\cup\lbrace c\rbrace)$. 

\emph{Case 2}: Let $|\lbrace a,b\rbrace\cap A_1|<2$ hold. If $c\in A_1$, then $u_1(A_1)\leq -3$ whereas $u_2(A_2)\geq -2$. But, then $u_1(A_1)<-2\leq u_2(A_2)=u_2(A_2\cup\lbrace c\rbrace)$ holds. If $c\in A_2$, then $u_1(A_1)\geq 0$ whereas $u_2(A_2)\leq -1$. Now, $u_2(A_2)<0=u_1(A_1 \setminus \lbrace o\rbrace)$  for each $o\in \lbrace a,b\rbrace\cap A_1$ and $u_2(A_2)<1\leq u_1(A_1\cup\lbrace o\rbrace)$ for each $o\in \lbrace a,b\rbrace\cap A_2$. 
\myqed
\end{myproof} 

This result compares favorably against an axiomatic property such as envy-freeness up to some item in the sense that EF1 allocations exist in each problem \cite{aziz2019gc}. In response to this axiomatic result, we study the following computational question related to JF1 allocations.

\vspace{0.5em}
\fbox{
\begin{minipage}{0.85\columnwidth}
 {\sc PossibleJF1(PossibleJF1andPO)} \\
{\em Data}: a division problem $([n],[m],(u_a(o))_{n\times m})$ \\
{\em Result}: is there an JF1 (JF1 and PO) allocation?
\end{minipage}
}
\vspace{0.5em}

We relate this problem to the well-known $\NP$-hard problem {\sc X3C} (i.e.\ exact cover by 3-sets) \cite{garey1979}. In fact, we observe that each instance of {\sc X3C} can be reduced in polynomial time to an instance of {\sc PossibleJF1}. We use this in our next result.

\vspace{0.5em}
\fbox{
\begin{minipage}{0.85\columnwidth}
 {\sc X3C} {\em Data}: a set $X=\lbrace x_1,\ldots,x_{3q}\rbrace$ for some $q\in\mathbb{N}_{\geq 2}$ and a collection $\mathcal{C}=\lbrace C| C\subseteq X, |C|=3\rbrace$ \\
{\em Result}: is there $\mathcal{C}^{\prime}\subseteq\mathcal{C}$ s.t. $\cup_{C\in\mathcal{C}^{\prime}}=X$?
\end{minipage}
}
\vspace{0.5em}

\begin{mytheorem}\label{thm:hardmixed}
In fair division of mixed manna and normalised additive utilities, {\sc PossibleJF1} is $\NP$-hard.
\end{mytheorem}  

\begin{myproof}
The {\sc X3C} problem is $\NP$-hard whenever $|\mathcal{C}|>q$. It remains $\NP$-hard whenever $|\mathcal{C}|>3q$. To see this, we note that each instance with at most $3q$ $3$-sets can be reduced to an instance with strictly more than $3q$ $3$-sets by simply copying a given $3$-set $3q-|\mathcal{C}|+1$ times (i.e.\ $\geq 1$ time). For this reason, we assume that $|\mathcal{C}|>3q$ holds.

We next present the polynomial-time reduction from  {\sc X3C} to {\sc PossibleJF1}. Let $X=\lbrace x_1,\ldots,x_{3q}\rbrace$, $q\geq 2$ and $\mathcal{C}=\lbrace C_1,\ldots,C_Q\rbrace$. The set of agents is $[Q+1]$. The set of items is $\lbrace x_1,\ldots,x_{3q},$ $y^1_1,y^2_1,y^3_1\ldots,y^1_{Q-q+1},y^2_{Q-q+1},y^3_{Q-q+1},$ $z\rbrace$. Let $M\in\mathbb{N}_{\geq 0}$ be such that $M\geq 3Q-3q+7$ holds. The utilities of agents for items are:

\begin{itemize}
\item for agent $a\in [Q]$: $u_a(x_i)$ is $1$ if $x_i\in C_a$ and else $-M$, $u_a(y^j_k)=1$ and $u_a(z)=0$,
\item for agent $(Q+1)$: $u_{Q+1}(x_i)=0$, $u_{Q+1}(y^j_k)=1$ and $u_{Q+1}(z)=-(3q-3)M+3$.
\end{itemize}

We note that the agents' utilities for the set of items are normalised and sum up to $-(3q-3)M+3+3(Q-q+1)$. We next prove that there is an exact cover by $3$-sets in the instance of {\sc X3C} iff there is a JF1 allocation in the instance of {\sc PossibleJF1}. 

Let $\mathcal{C}^{\prime}$ be an exact cover for $X$. Hence, $|\mathcal{C}^{\prime}|=q$ and $C_a\cap C_b=\emptyset$ for each $C_a,C_b\in\mathcal{C}^{\prime}$ with $C_a\neq C_b$. Wlog, let $\mathcal{C}^{\prime}=\lbrace C_1,\ldots,C_q\rbrace$. We construct the following allocation: $A_a=C_a$ for $a\in \lbrace 1,\ldots,q\rbrace$, $A_b=\lbrace y^1_{b-q},y^2_{b-q},y^3_{b-q}\rbrace$ for $b\in \lbrace q+1,\ldots,Q,Q+1\rbrace$. Wlog, give item $z$ to agent 1 because $u_1(z)=0$: $A_1=A_1\cup\lbrace z\rbrace$. Each agent receives utility of exactly $3$ in $A$. Hence, $A$ satisfies JF1.

Let $A$ be a JF1 allocation. We let $N$ denote the set of agents who receive the $x$s. Wlog, $N=\lbrace a_1,\dots,a_l\rbrace$. We note that $l\leq 3q$ because the number of $x$s is $3q$. Hence, each agent from $[Q+1]\setminus N$ does not receive any item from the $x$s. We note that there are at least two agents in $[Q+1]\setminus N$ because $Q\geq 3q+1$ and, therefore, $Q+1-l\geq Q+1-3q\geq 2$ hold. Hence, there is an agent from $[Q]$, say $a$, such that $u_a(A_a)\geq 0$. 


Let us assume that $z\in A_{Q+1}$. Therefore, $u_{Q+1}(A_{Q+1})\leq -M$ holds by the choices of $M$ and $q$ even if agent $(Q+1)$ receive all items. But, then we have $u_{Q+1}(A_{Q+1})<0\leq u_a(A_a)$. For each $o\in A_a$ with $u_a(o)>0$, it follows $u_{Q+1}(A_{Q+1})<0\leq u_a(A_a\setminus\lbrace o\rbrace)$ because of $A_a\cap\lbrace x_1,\ldots,x_{3q}\rbrace=\emptyset$. Further, $u_{Q+1}(A_{Q+1})<0\leq u_a(A_a\cup\lbrace z\rbrace)=u_a(A_a)$ because of $u_a(z)=0$. We note that $z$ is the only bad in $A_{Q+1}$. Hence, $A$ cannot satisfy JF1.

We conclude $z\not\in A_{Q+1}$ and $u_{Q+1}(A_{Q+1})\geq 0$. Let us next assume that some agent $b\in [Q]$ receive at least one item among the $x$s valued with $-M$. By the choice of $M$, it follows that $u_b(A_b)\leq 3Q-3q+6-M\leq -1$ holds. This implies $u_b(A_b)<u_{Q+1}(A_{Q+1})$. Moreover, we note that each $o$ in $A_b$ with $u_b(o)=-M$ is such that $u_{Q+1}(o)=0$. Therefore, $u_b(A_b)<0\leq u_{Q+1}(A_{Q+1}\cup\lbrace o\rbrace)$. For each pure good $o$ in $A_{Q+1}$, it also follows $u_b(A_b)<0\leq u_{Q+1}(A_{Q+1}\setminus\lbrace o\rbrace)$. We derive again a contradiction with the JF1 of $A$.

We conclude that each agent $b\in [Q]$ receives $x$s valued with $1$. It follows that the sum of agents' utilities in $A$ is $3Q+3$. If $u_c(A_c)\leq 2$ for some $c\in [Q]$, then $u_d(A_d)\geq 4$ for some $d\in [Q+1]$ with $d\neq c$. In this case, $u_c(A_c)<u_d(A_d)$. Also, $u_c(A_c)\leq 2<3=u_d(A_d\setminus\lbrace o\rbrace)$ holds for each $o\in A_d$ with $u_d(o)=1$. If $u_c(A_c)\geq 4$ for some $c\in [Q]$, then $u_d(A_d)\leq 2$ for some $d\in [Q+1]$ with $d\neq c$. Hence, $A$ falsifies again JF1. Therefore, $u_e(A_e)=3$ for each $e\in [Q]$ and $u_{Q+1}(A_{Q+1})=3$. 

As the number of $x$s and $y$s is $3(Q+1)$, it follows that each $e\in [Q]$ receives $3$ of these items. Also, agent $Q+1$ receive $3$ of the $y$s because they have zero utilities for the $x$s. Consequently, $q$ agents from $[Q]$ receive the $x$s and $Q-q+1$ agents from $[Q+1]$ receive the $y$s in $A$. This implies that $l=q$, $N=\lbrace a_1,\ldots,a_q\rbrace$ and $Q+1\not\in N$ hold. We note that $A_f\cap A_g=\emptyset$ holds for each $f,g\in N$ with $f\neq g$. We can now construct an exact cover by $3$-sets for $X$ by uniting the bundles of the agents in $N$: $\cup_{h\in N} A_h=X$. 
\myqed
\end{myproof} 

It follows by this result that checking whether JF1$_0$, JFX or JFX$_0$ (and PO) allocations exist is intractable. Indeed, supposing that there is a polynomial-time algorithm for such allocations would lead to a contradiction with our complexity result unless $\P=\NP$.

\section{JFX$_0$ and EFX$_0$ without mixed items}\label{sec:imp}

The presence of mixed items in the problem may make it im-possible to achieve even the weakest concept JF1. By comparison, the strongest concepts JFX$_0$ and EFX$_0$ might be violated by any allocation in the problem even if we remove the mixed items. This follows because some moved items in some allocations are valued with zero marginal utilities.

\begin{myproposition}\label{pro:impefxjfxzero}
There are problems with \num{2} agents and normalised additive utilities for \num{1} pure good and \num{2} bads, in which \emph{no} allocation is JFX$_0$ or EFX$_0$.
\end{myproposition}  

\begin{myproof}
Let us consider \num{2} agents, \num{1} pure good and \num{2} bads. We let both agents like the good but dislike different bads. 

\begin{center}
\begin{tabular}{|c|c|c|c|} \hline
   & a & b & c \\ \hline
   agent 1 & $2$ & $-1$ & $0$ \\
  agent 2 & $2$ & $0$ & $-1$ \\ \hline
\end{tabular}
\end{center}

By the symmetry of the agents' utilities for item $a$, we consider only four allocations: $A=(\lbrace a,b\rbrace,\lbrace c\rbrace)$, $B=(\lbrace a,c\rbrace,\lbrace b\rbrace)$, $C=(\lbrace b,c\rbrace,\lbrace a\rbrace)$, $D=(\lbrace a,b,c\rbrace,\emptyset)$. 

We argue that none of these allocations is JFX$_0$ or EFX$_0$. To see this for EFX$_0$, we give one violation of this property for each allocation: (1) $u_2(A_2\setminus\lbrace c\rbrace)=0<2=u_2(A_1)$, (2) $u_2(B_2\setminus\lbrace b\rbrace)=0<1=u_2(B_1)$, (3) $u_1(C_1\setminus\lbrace b\rbrace)=0<2=u_1(C_2)$ and (4) $u_2(D_2)=0<1=u_2(D_1\setminus\lbrace b\rbrace)$. 

We next give all violations of JFX$_0$  for each allocation: (1) $u_2(A_2)=-1<1=u_1(A_1\cup\lbrace c\rbrace)$, (2) $u_2(B_2)=0<2=u_1(B_1\setminus\lbrace c\rbrace)$, $u_2(B_2)=0<1=u_1(B_1\cup\lbrace b\rbrace)$ (3) $u_1(C_1)=-1<2=u_2(C_2\cup\lbrace b\rbrace)$, $u_1(C_1)=-1<1=u_2(C_2\cup\lbrace c\rbrace)$ and (4) $u_2(D_2)=0<1=u_1(D_1\setminus\lbrace c\rbrace)$.\myqed
\end{myproof}

This result diverges from existing results. For example, EFX$_0$ allocations exist in problems with \num{2} agents and goods \cite{plaut2018}. Also, JFX$_0$ allocations exist in problems with any number of agents and pure goods \cite{gourves2014}. It follows that adding bads to such problems breaks these results. 

\section{JFX without mixed items}\label{sec:jfx}

JFX$_0$ coincides with JFX whenever the problem contains pure goods and pure bads. This is not true in problems with goods and bads. In this case, JFX$_0$ allocations may not exist whereas JFX allocations are guaranteed to exist. For example, the leximin$++$ solution is JFX. 

\begin{mytheorem}\label{thm:jfxlex}
In fair division of goods and bads with general utilities, the leximin$++$ solution satisfies JFX.
\end{mytheorem}  

\begin{myproof}
Let $A$ be an leximin$++$ allocation. Suppose that $A$ is not JFX for a pair of agents $a,b\in [n]$ with $a\neq b$. That is, $u_a(A_a)<u_b(A_b)$. Also, (1) $u_a(A_a)<u_b(A_b\cup\lbrace o\rbrace)$ holds for some $o\in A_a$ with $u_a(A_a)<u_a(A_a\setminus\lbrace o\rbrace)$ or (2) $u_a(A_a)<u_b(A_b\setminus \lbrace o\rbrace)$ holds for some $o\in A_b$ with $u_b(A_b)>u_b(A_b\setminus\lbrace o\rbrace)$. Wlog, let $u_1(A_1)\leq\ldots\leq u_n(A_n)$ denote the utility order induced by $A$. We let $k=\argmax \lbrace i\in [n]|u_i(A_i)\leq u_a(A_a)\rbrace$. We note $a\leq k$ and $k<b$. We consider two cases. 

\emph{Case 1}: Let (1) hold for bad $o\in A_a$. Let us move item $o$ from the bundle $A_a$ to the bundle $A_b$. We let $C$ denote this new allocation. That is, $C_a=A_a\setminus\lbrace o\rbrace$, $C_b=A_b\cup\lbrace o\rbrace$ and $C_c=A_c$ for each $c\in [n]\setminus\lbrace a,b\rbrace$. We argue that $C\succ_{++} A$ holds. 

We note that $C_c=A_c$ holds for each $c\in [k]\setminus\lbrace a\rbrace$. We show $u_{a_k}(C_{a_k})>u_k(A_k)$ where $a_k$ is the $k$th agent in the utility order induced by $C$. If this agent is $a$, then $u_a(C_a)=u_a(A_a\setminus\lbrace o\rbrace)>u_a(A_a)=u_k(A_k)$ by (1). If this agent is $b$, then $u_b(C_b)=u_b(A_b\cup \lbrace o\rbrace)>u_a(A_a)=u_k(A_k)$ by (1). Otherwise, $u_{a_k}(C_{a_k})=u_{k+1}(A_{k+1})>u_k(A_k)$ by the definition of $k$. It follows in each case that $u_d(C_d)\geq u_{a_k}(C_{a_k})>u_k(A_k)$ holds for each agent $d\in [n]\setminus[([k]\setminus\lbrace a\rbrace)\cup\lbrace a_k\rbrace]$. Therefore, $A$ cannot be leximin$++$. This is a contradiction.

\emph{Case 2}: Let (2) hold for good $o\in A_b$. Let us move only item $o$ from $A_b$ to $A_a$. We let $B$ denote this allocation: $B_a=A_a\cup\lbrace o\rbrace$, $B_b=A_b\setminus\lbrace o\rbrace$ and $B_c=A_c$ for each $c\in [n]\setminus\lbrace a,b\rbrace$. Similarly as for $C$ in the first case, we next argue that $B\succ_{++} A$ holds. 

As item $o$ is good, it follows $u_a(B_a)\geq u_a(A_a)$. If $u_a(B_a)=u_a(A_a)$, then $B_c=A_c$ holds for each $c<a$, $|B_a|=|A_a|+1$ and $B_d=A_d$ for each $d\in (a,k]$. Moreover, it follows that $u_e(C_e)>u_k(A_k)=u_a(A_a)$ holds for each agent $e\in [n]\setminus [k]$, including for agent $b$ by (2). As $\succ_{++}$ maximizes the bundle size as a secondary objective, it follows that $B$ is strictly larger than $A$ under $\succ_{++}$. Again, $A$ cannot be leximin$++$. If $u_a(B_a)>u_a(A_a)$, then we reach a contradiction in a similar way as in the first case.
\myqed
\end{myproof}

This characterization result is tight. Indeed, if we relaxed JFX to JFX$^0$ in it, then we would derive an impossibility result by Proposition~\ref{pro:impefxjfxzero}. 

\section{JF1$_0$ without mixed items}\label{sec:jfonegb}

The leximin$++$ solution is JFX and it can be computed in $O(n^m)$ time. This might be fine for small $m$. However, $m$ can be much larger than $n$ in practice. For this reason, we may wish to return an allocation that satisfies the weaker concept JF1$_0$. Surprisingly, we can do this in $O(mn)$ time by using Algorithm~\ref{alg:jfone}. We next describe this algorithm.

Basically, Algorithm~\ref{alg:jfone} allocates the items one-by-one to agents in an arbitrary order. If the current item is a pure good, then it goes to an agent with minimum utility. If it is a pure bad, then it goes to an agent with maximum utility, supposing the item is given to them. Otherwise, it goes to an agent who has zero utility for it. 

The key idea behind the inductive proof of the correctness of the algorithm relies on the fact that it differentiates between pure goods, pure bads and indifferent items. Thus, it is guaranteed that if the partial allocation at a given round were not JF1$_0$, then the partial allocation at the previous round would also violate JF1$_0$.

\begin{algorithm}
\caption{Jealousy-freeness up to some item}\label{alg:jfone}
\begin{algorithmic}[1]
\Procedure{JF1$_0$Allocation}{$[n],[m],(u_a)_{n}$} 
\State $\forall a\in [n]: A_a\gets\emptyset$
\For{$t=1:m$}
\If{$\forall c\in [n]: u_c(A_c\cup\lbrace t\rbrace)>u_c(A_c)$} 
\State $a\gets\arg\min_{b\in [n]} u_b(A_b)$
\ElsIf{$\forall c\in [n]: u_c(A_c\cup\lbrace t\rbrace)<u_c(A_c)$}
\State $a\gets\arg\max_{b\in [n]} u_b(A_b\cup\lbrace t\rbrace)$
\Else 
\State $a\gets\arg\lbrace b\in [n]|u_b(A_b\cup\lbrace t\rbrace)=u_b(A_b)\rbrace$
\EndIf
\State $A_a\gets A_a\cup\lbrace t\rbrace$
\EndFor
\State \Return $A$ 
\EndProcedure
\end{algorithmic}
\end{algorithm}

\begin{mytheorem}\label{thm:jfonealg}
In fair division of goods and bads with general utilities, Algorithm~\ref{alg:jfone} returns an JF1$_0$ allocation.
\end{mytheorem}

\begin{myproof} 
Algorithm~\ref{alg:jfone} gives the items in rounds $1$ to $m$. We let $A^t$ denote the allocation constructed up to round $t$. We will prove that each $A^t$ is JF1$_0$ by induction on $t$. 

In the base case when $t=1$, the proof is trivial. In the hypothesis, let us assume that $A^{t-1}$ is JF1$_0$. In the step case, Algorithm~\ref{alg:jfone} allocates item $t$. Wlog, let agent 1 receive $t$. Thus, the bundle of each other agent in $A^t$ is the same as in $A^{t-1}$. Hence, each two agents $a,b\in [n]\setminus\lbrace 1\rbrace$ with $a\neq b$ are JF1$_0$ of each other in $A^t$ by the hypothesis. For this reason, we next consider three remaining cases.

\emph{Case 1}: If $u_1(A^t)=u_1(A^{t-1})$, then agent 1's utility in $A^t$ is equal to their utility in $A^{t-1}$. Hence, the allocation $A^t$ remains JF1$_0$.  

\emph{Case 2}: If $u_1(A^t)<u_1(A^{t-1})$, then $t$ must be a pure bad. By the hypothesis, it follows that each other agent is JF1$_0$ of agent 1 in $A^t$ simply because 1's utility in $A^t$ is strictly lower than 1's utility in $A^{t-1}$. Consequently, we only prove that agent 1 is JF1$_0$ of each agent $a\in [n]\setminus\lbrace 1\rbrace$. If this were not the case for some $a\in [n]\setminus\lbrace 1\rbrace$, then $u_1(A^t_1)<u_a(A^t_a\cup \lbrace t\rbrace)$ would hold. This would imply $u_1(A^{t-1}_1\cup \lbrace t\rbrace)<u_a(A^{t-1}_a\cup \lbrace t\rbrace)$ because of $A^t_1=A^{t-1}_1\cup \lbrace t\rbrace, A^t_a=A^{t-1}_a$ and contradict the fact that $u_1(A^{t-1}_1\cup \lbrace t\rbrace)\geq u_a(A^{t-1}_a\cup \lbrace t\rbrace)$ holds.

\emph{Case 3}: If $u_1(A^t)>u_1(A^{t-1})$, then $t$ must be a pure good. As agent 1 is JF1$_0$ of any other agent in $A^{t-1}$, they remain JF1$_0$ of them in $A^t$ simply because 1's utility in $A^t$ is strictly greater than 1's utility in $A^{t-1}$. Hence, we only prove that each agent $a\in [n]\setminus\lbrace 1\rbrace$ is JF1$_0$ of agent 1. As agent 1 gets item $t$, it must be that $u_a(A^{t-1}_a)\geq u_1(A^{t-1}_1)$ holds. Also, $u_a(A^t_a)=u_a(A^{t-1}_a)$ and $u_1(A^{t-1}_1)=u_1(A^t_1\setminus\lbrace t\rbrace)$. Hence, $u_a(A^t_a)\geq u_1(A^t_1\setminus \lbrace t\rbrace)$. 
\myqed
\end{myproof}

We further consider combinations of jealousy-freeness up to one item and Pareto-optimality.

\section{JF1, JFX and PO}\label{sec:jfonepo}

Freeman et al.\ \shortcite{freeman2019ijcai} gave a problem with \num{3} agents and normalised additive and binary (i.e.\ $0/1$) utilities, where none of the allocations satisfies simultaneously JF1$_0$ and PO. This incompatibility might further hold in our setting even with just \num{2} agents.

\begin{myproposition}\label{pro:impall}
There are problems with \num{2} agents and normalised additive utilities for \num{2} goods and \num{2} bads, where \emph{no} PO allocation is JF1.
\end{myproposition}  

\begin{myproof}
Let us consider the below problem with \num{2} agents, \num{2} goods and \num{2} bads. We note that the agents' utilities are normalised.

\begin{center}
\begin{tabular}{|c|c|c|c|c|} \hline
   & a & b & c & d  \\ \hline
  agent 1 & $1$ & $1$ & $-5$ & $0$ \\
  agent 2 & $0$ & $0$ & $0$ & $-3$  \\ \hline
\end{tabular}
\end{center}

There is only one PO allocation (leximin). This one gives items $a$, $b$, $d$ to agent 1 and item $c$ to agent 2. Let $A$ denote this allocation. We have $u_1(A_1)=2$, $u_2(A_2)=0$ and $u_2(A_2)<1=u_1(A_1\setminus\lbrace o\rbrace)$ for each $o\in A_1$ with $u_1(o)>0$. Hence, the allocation $A$ violates JF1.
\myqed
\end{myproof}  

This result reveals the technical difference between the leximin and leximin$++$ solutions. The former one is PO but may not be JF1 because it could give to an agent high utility whilst the latter one is JFX but may not be PO because it could give to an agent an item for which they have zero utility.

Freeman et al.\ \shortcite{freeman2019ijcai} proved that deciding whether JF1$_0$ and PO allocations exist in problems with non-normalised and non-binary utilities is $\NP$-hard. We strengthen this hardness result to the weaker combination of JF1 and PO in problems with normalised and binary utilities.

\begin{mytheorem}\label{thm:hardgoods}
In fair division of goods and normalised additive and $0/1$ utilities, {\sc PossibleJF1andPO} is $\NP$-hard.
\end{mytheorem} 

\begin{myproof}
Let us consider the reduction in Theorem~\ref{thm:hardmixed}. Suppose that we remove agent $(Q+1)$ and items $y^1_{Q-q+1}$, $y^2_{Q-q+1}$, $y^3_{Q-q+1}$, $z$. Further, suppose that we substitute each $-M$ with $0$. This transformation gives us a problem with $Q$ agents, $3Q$ goods and normalised $0/1$ utilities. Let $\mathcal{C}$ denote an exact cover for $X$. We can construct an allocation $A_C$ as in the proof of Theorem~\ref{thm:hardmixed}. It follows that $A_C$ is JF1. This allocation is also PO because each agent receive items valued with $1$. Let there be an JF1 and PO allocation $A$. By PO, it must be the case that each agent receive items valued with $1$. Hence, the sum of agents' utilities in $A$ is equal to $3Q$. This is only possible whenever each agent get utility $3$. We can construct an exact cover $C_A$ for $X$ as in the proof of Theorem~\ref{thm:hardmixed}. The result follows.
\myqed
\end{myproof}

The impossibility result differs from the existence of EF1 and PO allocations in the case of $0/1$ utilities. In fact, such allocations can be computed in polynomial time by the online algorithm {\sc Balanced Like} from \cite{aleksandrov2015ijcai}. Benade et al.\ \shortcite{benade2018} made this observation. 

The result further breaks whenever there are just pure items in the problem (i.e.\ non-zero marginal utilities). In fact, an allocation in such a problem is leximin$++$ iff it is leximin. By Remark~\ref{rem:one} and Theorem~\ref{thm:jfxlex}, it follows that the leximin solution is JFX and the leximin$++$ solution is PO.

We can safely extend these guarantees to problems where agents are indifferent for some bads. Indeed, the leximin solution is PO and, for this reason, it gives each such bad to some agent who has zero marginal utility for it. These decisions are optimal from a JFX perspective as well.

\begin{mycorollary}\label{cor:jfxone}
In fair division of pure goods and bads with general utilities, the leximin solution satisfies JFX and PO.
\end{mycorollary}  

This result is tight by Proposition~\ref{pro:impefxjfxzero}. This also follows by an existing result of Freeman et al.\ \shortcite{freeman2020equitable} who observed that JFX$_0$ and PO might not be attainable in problems with bads.

\section{JF1, JFX and EF1, EFX}\label{sec:jfoneefone} 

JF1 and EF1 might be unachievable in problems where some agents have zero total utility for the items (see Example~\ref{exp:normtwo}). For this reason, we assume in this section that there are no such agents. We present indeed some positive results under this common assumption.

\subsection{The case of $2$ agents}\label{subsec:two}

JF1 and EF1 might as well be violated by each allocation in problems with non-normalised utilities for bads. This is because JF1 may bias the allocation towards agents with the greatest total utility for bads. Thus, such an allocation may give all bads to a single agent. As a result, it could easily falsify an axiomatic property such as EF1.

\begin{myproposition}\label{pro:impnorm}
There are problems with \num{2} agents and additive utilities for \num{2} pure bads, where \emph{no} JF1 allocation is EF1.
\end{myproposition}  

\begin{myproof}
The proof is in terms of the below counter-example. The agents' utilities are clearly not normalised, one with a total of $-2$ and the other one with a total of $-6$.

\begin{center}
\begin{tabular}{|c|c|c|} \hline
   & a & b    \\ \hline
 agent 1 & $-1$ & $-1$    \\ 
 agent 2 & $-3$ & $-3$   \\ \hline
\end{tabular}
\end{center}

To achieve JF1, we argue that we should give both items to agent 1. Otherwise, agent 2 would get disutility of at least $-3$ but be still jealous up to one item of agent 1 because shifting one item from 2's bundle to 1's bundle would make 1's disutility at most $-2$. Well, let us then give the items to agent 1. Clearly, this violates EF1 because agent 1 envies agent 2 even after removing any item from their 1's bundle.
\myqed
\end{myproof}

It follows that the leximin and leximin$++$ solutions might violate EF1 in some problems with non-normalised utilities. At the same time, it seems to us that normalisation occurs often in practice. For example, some web-applications on Spliddit ask agents to share a fixed total (i.e.\ normalised) utility for items \cite{caragiannis2016}.

As a response, we will prove shortly that the impossibility result breaks in two contexts with normalised additive utilities: (1) JFX, EFX and PO for pure goods and bads; (2) JFX and EFX for goods and bads. However, we first give another strong result. Namely, EFX and PO are always attainable in problems with such utilities for arbitrary items.

\begin{mytheorem}\label{thm:efonetwo}
In fair division of mixed manna with \num{2} agents and normalised additive utilities, the leximin solution satisfies EFX and PO.
\end{mytheorem}  

\begin{myproof}
Let $A$ be an leximin allocation. By Remark~\ref{rem:one}, $A$ is PO. Suppose that $A$ is not EFX. Wlog, let agent 1 be not EFX of agent 2. Hence, it must be the case that (1) $u_1(A_1\setminus\lbrace o\rbrace)<u_1(A_2)$ holds for some $o\in A_1$ with $u_1(o)<0$ or (2) $u_1(A_1)<u_1(A_2\setminus\lbrace o\rbrace)$ holds for some $o\in A_2$ with $u_1(o)>0$. We consider two cases.

If (1) holds for $o\in A_1$ with $u_1(o)<0$, then $u_2(o)<0$ by the PO of $A$. Let us consider bundles $S_1=A_1\setminus\lbrace o\rbrace$ and $S_2=A_2\cup\lbrace o\rbrace$ in this case. 

If (2) holds for $o\in A_2$ with $u_1(o)>0$, then $u_2(o)>0$ by the PO of $A$. Let us consider bundles $S_1=A_1\cup\lbrace o\rbrace$ and $S_2=A_2\setminus\lbrace o\rbrace$ in this case. 

We construct an allocation $B$ and show that the minimum utility in $B$ is greater than the minimum utility in $A$, reaching a contradiction with the leximin-optimality of $A$. We let 

\begin{equation*}
B_1=\argmin_{S\in\lbrace S_1,S_2\rbrace} u_2(S),
\end{equation*}
\begin{equation*}
B_2=\argmax_{S\in\lbrace S_1,S_2\rbrace} u_2(S).
\end{equation*}

By construction, $u_2(B_2)\geq u_2(B_1)$ holds in $B$. Moreover, $u_1(S_1)>u_1(A_1)$ and $u_1(S_2)>u_1(A_1)$ hold in each of the cases (1) and (2). These inequalities follow because agent 1's utilities for the moved item are non-zero and agent 1 is not EFX of agent 2. We conclude $u_1(B_1)>u_1(A_1)$.

We have $u_1(A_1)+u_1(A_2)=c$ and $u_2(A_1)+u_2(A_2)=c$ for some $c\in \mathbb{R}$ by the fact that the agents' utilities are normalised and additive. As $u_1(A_1)<u_1(A_2)$, it follows $u_1(A_1)<c/2$. By the PO of $A$, $u_2(A_2)>u_2(A_1)$. Hence, $u_2(A_1)<c/2$ and $u_2(A_2)>c/2$. Further, as $u_2(B_2)\geq u_2(B_1)$ and $u_2(B_1)+u_2(B_2)$ $=c$, it follows $u_2(B_2)\geq c/2$. 

We are ready to derive the aforementioned contradiction: $\min\lbrace u_1(A_1),u_2(A_2)\rbrace=u_1(A_1)<\min\lbrace u_1(B_1),c/2\rbrace$ $\leq\min\lbrace u_1(B_1),u_2(B_2)\rbrace$. These follow because of $u_2(A_2)>c/2$, $u_1(A_1)<c/2$, $u_1(A_1)<u_1(B_1)$ and $u_2(B_2)\geq c/2$. The result follows.
\myqed
\end{myproof} 

This result provides stronger guarantees than some existing results. For example, the MNW solution is guaranteed to be EF1 and PO in problems with goods \cite{caragiannis2016}. On the other hand, it may violate EF1 or PO in problems with mixed items (see \cite{aleksandrov2019greedy}) where the leximin solution remains EFX and PO.

Theorem~\ref{thm:efonetwo} holds for problems with normalised additive utilities. By comparison, EFX and PO may be incompatible in problems with normalised general utilities for pure goods (i.e.\ with non-zero marginal utilities for goods) \cite{plaut2018}. As a consequence, Theorem~\ref{thm:efonetwo} breaks with general utilities.

By Corollary~\ref{cor:jfxone}, the leximin solution is JFX in problems with normalised additive utilities for pure goods and bads. Furthermore, both the leximin and leximin$++$ solutions induce now the same utilities but their distributions of indifferent bads may differ. By Theorem~\ref{thm:efonetwo}, the leximin$++$ solution is EFX and PO in such problems. 

\begin{mycorollary}\label{cor:jfxtwo}
In fair division of pure goods and bads with 2 agents and normalised additive utilities, the leximin solution satisfies JFX, EFX and PO.
\end{mycorollary}  

By Proposition~\ref{pro:impall}, we cannot extend this result to problems where the agents are indifferent for goods. However, Plaut and Roughgarden \shortcite{plaut2018} argued that each such good (or even a mixed item in our view) could be given to the agent who values it positively prior to the allocation.

Otherwise, Proposition~\ref{pro:impall} implies that no PO allocation can satisfy both EF1 and JF1, including the leximin solution. Nevertheless, we might wish to drop PO and achieve only EFX and JFX. Surprisingly, we can do this in a special case. For this purpose, we now need the leximin$++$ solution.

\begin{mytheorem}\label{thm:efxjfx}
In fair division of goods and bads with 2 agents and normalised additive utilities, the leximin$++$ solution is JFX and EFX.
\end{mytheorem}  

\begin{myproof}
Let $A$ denote an leximin$++$ allocation. By Theorem~\ref{thm:jfxlex}, JFX follows. We next prove EFX. Suppose that agent 1 is not EFX of agent 2. Consequently, $u_1(A_1)<u_1(A_2)$. Further, as the utilities of agent 1 are normalised, it follows that $u_1(A_1)<c/2$ holds for some $c\in\mathbb{R}$. We construct a new allocation $B$ as in Theorem~\ref{thm:efonetwo}. 

We have $u_1(B_1)>u_1(A_1)$. We also show that $u_2(B_2)>u_1(A_1)$ holds. The argument for it is by contradiction. Let $u_2(B_2)\leq u_1(A_1)$ hold. By the definition of $B$, $u_2(B_2)\geq u_2(B_1)$. Therefore, $u_2(B_2)<c/2$ and $u_2(B_1)<c/2$ follow because $u_1(A_1)<c/2$. This is in conflict with $u_2(B_1)+u_2(B_2)=c$. Hence, $u_2(B_2)>u_1(A_1)$. 

Let $k=\min\lbrace u_1(A_1),u_2(A_2)\rbrace$. We derive $u_1(B_1)>k$ and $u_2(B_2)>k$. The minimum utility in $B$ is strictly greater than the minimum utility in $A$. As an leximin$++$ allocation maximizes this utility before the bundle size, it follows that $B$ is strictly larger than $A$ under $\succ_{++}$. This means that $A$ cannot be leximin$++$. We reached a contradiction.
\myqed
\end{myproof}

The leximin and leximin$++$ solutions are both intractable. We can however compute in $O(mn)$ time an EFX allocation in problems with additive utilities. For this purpose, we can use the ``cut-and-choose'' protocol from \cite{plaut2018} with the EFX algorithm for identical utilities from \cite{aleksandrov2019greedy} as a sub-routine. 

With general utilities, there are problems where no allocation satisfies both JF1 and EF1, even under the assumption that the agents' utilities for the set of items are the same. Hence, Theorem~\ref{thm:efxjfx} breaks. That is, the leximin$++$ or even leximin solution may no longer satisfy even EF1 as we move to normalised general problems with pure goods.

\begin{myproposition}\label{pro:impjfxefx}
There are problems with \num{2} agents and normalised general utilities for \num{4} pure goods, where \emph{no} JF1 allocation is EF1.
\end{myproposition} 

\begin{myproof}
Let us consider \num{2} agents and a set of goods $[m]$, where $m\geq 4$. Define each agent's utilities as follows: (1) $u_1(S)=|S|$ for each $S\subseteq [m]$ and (2) $u_2(S)=\epsilon|S|$ for each $S$ such that $S\subset [m]$ and $u_2([m])=m$, where $\epsilon\in (0,\frac{1}{m})$. We note that the agents' utilities are normalised: $u_1(\emptyset)=u_2(\emptyset)=0$ and $u_1([m])=u_2([m])=m$. 

If agent 1 received at least two items, then their utility would be at least $2$. But, then agent 2 get at most $(m-2)$ items and, hence, their utility is at most $\epsilon(m-2)$. This is strictly lower than $1$ because of $\epsilon<\frac{1}{m}$. As a result, agent 2 is jealous of agent 1 even after the removal of any item from 1's bundle. Hence, such an allocation cannot be JF1.

To achieve JF1, agent 1 should receive at most one item. If they received no item, then they would not be JF1 of agent 2. Hence, they receive one item and their utility is $1$. However, agent 2 now receive $(m-1)$ items and agent 1's utility for 2's bundle without one item is $(m-2)$. This is at least $2$ because of $m\geq 4$. Hence, such an allocation cannot be EF1.
\myqed
\end{myproof}

By Proposition~\ref{pro:impefxjfxzero}, the possibility results break whenever we relax JFX to JFX$_0$ or EFX to EFX$_0$. Hence, they are tight. This is also in-line with an impossibility result for EFX$^0$ and PO allocations of goods \cite{plaut2018}. 

\subsection{The case of $n\geq 3$ or fewer items}\label{sub:three}

In the case of \num{2} agents, normalisation plays a crucial role in achieving JFX and EFX in problems with goods and bads. However, normalisation may not help us whenever there are more agents in the problem. This result complements an existing impossibility result for JF1$_0$ and EF1 allocations in problems with \num{3} agents and non-normalised utilities for \num{7} pure goods \cite{freeman2019ijcai}.

\begin{myproposition}\label{pro:imphouse}
There are problems with \num{3} agents and normalised additive utilities for \num{3} pure bads, where \emph{no} JF1 allocation is EF1.
\end{myproposition}  

\begin{myproof}
The proof uses a simple counter-problem with \num{3} agents whose utilities for \num{3} pure bads sum up to $-30$.

\begin{center}
\begin{tabular}{|c|c|c|c|} \hline
   & a & b & c   \\ \hline
  agent 1 & $-28$ & $-1$ & $-1$    \\  
  agent 2 & $-24$ & $-3$ & $-3$    \\ 
  agent 3 & $-16$ & $-7$ & $-7$  \\\hline
\end{tabular}
\end{center}

To achieve EF1, it is easy to see that each agent should get exactly one item. Otherwise, an agent with at least two items would not be EF1 of an agent with zero items. Hence, there are $6$ EF1 allocations. 

For each allocation of $a$, there are two symmetrical EF1 allocations giving $b$ and $c$ to different agents. For this reason, let us consider only $3$ EF1 allocations: $A=(\lbrace a\rbrace,\lbrace b\rbrace,\lbrace c\rbrace)$, $B=(\lbrace b\rbrace,\lbrace a\rbrace,\lbrace c\rbrace)$ and $C=(\lbrace c\rbrace,\lbrace b\rbrace,\lbrace a\rbrace)$. 

We simply show that each of these allocations violates JF1: (1) $u_1(A_1)=-28<-27=u_2(A_2\cup\lbrace a\rbrace)$, (2) $u_2(B_2)=-24<-23=u_3(B_3\cup\lbrace a\rbrace)$ and (3) $u_2(C_2)=-3<-2=u_1(C_1\cup\lbrace b\rbrace)$. The result follows.
\myqed
\end{myproof}

This result suggests that we might wish to achieve JF1$_0$ and EF1 in isolation from each other. For JF1$_0$ in problems without mixed items, we can use Algorithm~\ref{alg:jfone}. For EF1 in problems with any items, we can use the generalized envy-graph algorithm from \cite{aziz2019gc}.

The case of pure bads contrasts axiomatically and computationally with the case of pure goods and additive utilities. In this case, JFX, EFX and PO allocations exist. For example, the allocation that maximizes the sum of agents' utilities for at most one pure good satisfies these three properties. Such an allocation can be computed in $O(n^3)$ time by using a matching procedure such as the hungarian method \cite{kuhn1955}.

In contrast, let us consider a problem with normalised general utilities for at most $n$ goods \emph{or} bads. Giving the goods (bads) to different agents is JFX and EFX (EFX). Such an allocation can be computed in $O(n)$ time. We submit it as an \emph{open} question whether JFX or EFX allocations can be computed efficiently in the case of at most $n$ goods \emph{and} bads. 

\section{Discussion}\label{sec:disc}

We could attempt to tackle the impossibility results by relaxing further JF1 or EF1. For example, two natural approximations of EF1 for goods are EF2 from \cite{bilo2019} and PROP1 from \cite{conitzer2017}. We can construct though problems as in Propositions~\ref{pro:impjfxefx} and~\ref{pro:imphouse} where neither EF2 nor PROP1 is compatible with JF1. Alternatively, we can approximate JF1 by relaxing the ``up to one item'' constraints to ``up to two items''. Say, we call this property JF2. We can again give similar problems where JF2 is incompatible with EF1. These observations suggest that further relaxations of EF2 and JF2 also might not interact in some problems. 

Even more, Algorithm~\ref{alg:jfone} for JF1$_0$ allocations remind us of the popular envy-graph algorithm for EF1 allocations \cite{lipton2004}. Both algorithms allocate the items one-by-one in some order. However, the envy-graph algorithm lets agents exchange bundles of some previous items at each step. By comparison, a notable advantage of Algorithm~\ref{alg:jfone} is that it makes the decision for the current item without re-allocating any of the previous items or using information about any of the next items. Hence, it can be adapted to work in online environments such as the one in \cite{aleksandrov2015ijcai} by inputting the items to it one-by-one. 

\section{Conclusions}\label{sec:conc}

We considered a fair division setting where agents assign utilities to bundles of indivisible items in a mixed manna. For this model, we studied combinations of properties for concepts such as jealousy-freeness up to one item, envy-freeness up to one item and Pareto-optimality. We obtained many possibility and impossibility results for such combinations. We also studied computational tasks related to these combinations: some of them exhibit exponential-time algorithms (unless $\P=\NP$) and some others admit polynomial-time algorithms. We summarized all our results in Table~\ref{tab:results}.

Our work opens up many questions for future work. For example, how well perform the proposed solutions in simulations? Also, the leximin$++$ solution is unfortunately intractable even when the agents' utilities are specified in unary. For this reason, we believe that it is natural to ask whether there is a pseudo-polynomial-time algorithm for JFX allocations in this case? Further, the leximin$++$ solution is EFX in problems with goods and general but identical utilities. It is, therefore, also pertinent to ask if these guarantees extend to problems with mixed manna? 

\bibliographystyle{named}
\bibliography{bads}

\begin{thebibliography}{}

\bibitem[\protect\citeauthoryear{Aleksandrov and
  Walsh}{2019}]{aleksandrov2019greedy}
Martin Aleksandrov and Toby Walsh.
\newblock Greedy algorithms for fair division of mixed manna.
\newblock {\em CoRR}, abs/1911.11005, 2019.

\bibitem[\protect\citeauthoryear{Aleksandrov \bgroup \em et al.\egroup
  }{2015}]{aleksandrov2015ijcai}
Martin Aleksandrov, Haris Aziz, Serge Gaspers, and Toby Walsh.
\newblock Online fair division: analysing a food bank problem.
\newblock In {\em Proceedings of the Twenty-Fourth International Joint
  Conference on Artificial Intelligence, {IJCAI} 2015, Buenos Aires, Argentina,
  July 25-31, 2015}, pages 2540--2546, 2015.

\bibitem[\protect\citeauthoryear{Aleksandrov \bgroup \em et al.\egroup
  }{2019}]{aleksandrov2019epia}
Martin Aleksandrov, Cunjing Ge, and Toby Walsh.
\newblock Fair division minimizing inequality.
\newblock In {\em Progress in Artificial Intelligence, 19th {EPIA} Conference
  on Artificial Intelligence, {EPIA} 2019, Vila Real, Portugal, September 3-6,
  2019, Proceedings, Part {II}}, pages 593--605, 2019.

\bibitem[\protect\citeauthoryear{Aziz \bgroup \em et al.\egroup
  }{2019a}]{aziz2019gc}
Haris Aziz, Ioannis Caragiannis, Ayumi Igarashi, and Toby Walsh.
\newblock Fair allocation of indivisible goods and chores.
\newblock In {\em Proceedings of the Twenty-Eighth International Joint
  Conference on Artificial Intelligence, {IJCAI-19}}, pages 53--59.
  International Joint Conferences on Artificial Intelligence Organization, 7
  2019.

\bibitem[\protect\citeauthoryear{Aziz \bgroup \em et al.\egroup
  }{2019b}]{aziz2019popropone}
Haris Aziz, Herv{\'{e}} Moulin, and Fedor Sandomirskiy.
\newblock A polynomial-time algorithm for computing a {P}areto optimal and
  almost proportional allocation.
\newblock {\em CoRR}, abs/1909.00740, 2019.

\bibitem[\protect\citeauthoryear{Benade \bgroup \em et al.\egroup
  }{2018}]{benade2018}
Gerdus Benade, Aleksandr~M. Kazachkov, Ariel~D. Procaccia, and
  Christos-Alexandros Psomas.
\newblock How to make envy vanish over time.
\newblock In {\em Proceedings of the 2018 ACM Conference on Economics and
  Computation}, EC '18, pages 593--610, New York, NY, USA, 2018. ACM.

\bibitem[\protect\citeauthoryear{Bez\'{a}kov\'{a} and
  Dani}{2005}]{bezakova2005ind}
Ivona Bez\'{a}kov\'{a} and Varsha Dani.
\newblock Allocating indivisible goods.
\newblock {\em Association for Computing Machinery ({ACM}) SIGecom Exchanges},
  5(3):11–18, April 2005.

\bibitem[\protect\citeauthoryear{Bil{\`{o}} \bgroup \em et al.\egroup
  }{2019}]{bilo2019}
Vittorio Bil{\`{o}}, Ioannis Caragiannis, Michele Flammini, Ayumi Igarashi,
  Gianpiero Monaco, Dominik Peters, Cosimo Vinci, and William~S. Zwicker.
\newblock Almost envy-free allocations with connected bundles.
\newblock In Avrim Blum, editor, {\em 10th Innovations in Theoretical Computer
  Science Conference, {ITCS} 2019, January 10-12, 2019, San Diego, California,
  {USA}}, volume 124 of {\em LIPIcs}, pages 14:1--14:21. Schloss Dagstuhl -
  Leibniz-Zentrum f{\"{u}}r Informatik, 2019.

\bibitem[\protect\citeauthoryear{Bliem \bgroup \em et al.\egroup
  }{2016}]{bliem2016}
Bernhard Bliem, Robert Bredereck, and Rolf Niedermeier.
\newblock Complexity of efficient and envy-free resource allocation: Few
  agents, resources, or utility levels.
\newblock In {\em Proceedings of the Twenty-Fifth International Joint
  Conference on Artificial Intelligence, New York, NY, USA, 9-15 July 2016},
  pages 102--108, 2016.

\bibitem[\protect\citeauthoryear{Brams and Taylor}{1996}]{brams1996}
Steven~J. Brams and Alan~D. Taylor.
\newblock {\em Fair Division - from Cake-cutting to Dispute Resolution}.
\newblock Cambridge University Press, 1996.

\bibitem[\protect\citeauthoryear{Caragiannis \bgroup \em et al.\egroup
  }{2012}]{caragiannis2012}
Ioannis Caragiannis, Christos Kaklamanis, Panagiotis Kanellopoulos, and Maria
  Kyropoulou.
\newblock The efficiency of fair division.
\newblock {\em Theory of Computing Systems}, 50(4):589--610, May 2012.

\bibitem[\protect\citeauthoryear{Caragiannis \bgroup \em et al.\egroup
  }{2016}]{caragiannis2016}
Ioannis Caragiannis, David Kurokawa, Herv{\'{e}} Moulin, Ariel~D. Procaccia,
  Nisarg Shah, and Junxing Wang.
\newblock The unreasonable fairness of maximum {N}ash welfare.
\newblock In {\em Proceedings of the 2016 Association for Computing Machinery
  {ACM} Conference on {EC} '16, Maastricht, The Netherlands, July 24-28, 2016},
  pages 305--322, 2016.

\bibitem[\protect\citeauthoryear{Conitzer \bgroup \em et al.\egroup
  }{2017}]{conitzer2017}
Vincent Conitzer, Rupert Freeman, and Nisarg Shah.
\newblock Fair public decision making.
\newblock In {\em Proceedings of the 2017 ACM Conference on Economics and
  Computation}, EC '17, pages 629--646, New York, NY, USA, 2017. ACM.

\bibitem[\protect\citeauthoryear{Dobzinski and
  Vondr\'{a}k}{2013}]{dobzinski2013}
Shahar Dobzinski and Jan Vondr\'{a}k.
\newblock Communication complexity of combinatorial auctions with submodular
  valuations.
\newblock In {\em Proceedings of the Twenty-fourth Annual ACM-SIAM Symposium on
  Discrete Algorithms}, SODA '13, pages 1205--1215, Philadelphia, PA, USA,
  2013. Society for Industrial and Applied Mathematics.

\bibitem[\protect\citeauthoryear{Dubins and Spanier}{1961}]{dubins1961}
L.~E. Dubins and E.~H. Spanier.
\newblock How to cut a cake fairly.
\newblock {\em The American Mathematical Monthly}, 68(1P1):1--17, 1961.

\bibitem[\protect\citeauthoryear{Endriss}{2013}]{endriss2013}
Ulle Endriss.
\newblock Reduction of economic inequality in combinatorial domains.
\newblock In Maria~L. Gini, Onn Shehory, Takayuki Ito, and Catholijn~M. Jonker,
  editors, {\em International conference on Autonomous Agents and Multi-Agent
  Systems, {AAMAS} '13, Saint Paul, MN, USA, May 6-10, 2013}, pages 175--182.
  {IFAAMAS}, 2013.

\bibitem[\protect\citeauthoryear{Freeman \bgroup \em et al.\egroup
  }{2019}]{freeman2019ijcai}
Rupert Freeman, Sujoy Sikdar, Rohit Vaish, and Lirong Xia.
\newblock Equitable allocations of indivisible goods.
\newblock In {\em Proceedings of the Twenty-Eighth International Joint
  Conference on Artificial Intelligence, {IJCAI-19}}, pages 280--286.
  International Joint Conferences on Artificial Intelligence Organization,
  2019.

\bibitem[\protect\citeauthoryear{Freeman \bgroup \em et al.\egroup
  }{2020}]{freeman2020equitable}
Rupert Freeman, Sujoy Sikdar, Rohit Vaish, and Lirong Xia.
\newblock Equitable allocations of indivisible chores.
\newblock {\em CoRR}, abs/2002.11504, 2020.

\bibitem[\protect\citeauthoryear{Garey and Johnson}{1979}]{garey1979}
M.~R. Garey and David~S. Johnson.
\newblock {\em Computers and Intractability: {A} Guide to the Theory of
  NP-Completeness}.
\newblock W. H. Freeman, 1979.

\bibitem[\protect\citeauthoryear{{Gini}}{1912}]{gini1912}
Corrado {Gini}.
\newblock {\em {Variabilit{\`a} e mutabilit{\`a}}}.
\newblock C. Cuppini, Bologna, 1912.

\bibitem[\protect\citeauthoryear{Gourv{\`{e}}s \bgroup \em et al.\egroup
  }{2013a}]{gourves2013}
Laurent Gourv{\`{e}}s, J{\'{e}}r{\^{o}}me Monnot, and Lydia Tlilane.
\newblock A matroid approach to the worst case allocation of indivisible goods.
\newblock In Francesca Rossi, editor, {\em {IJCAI} 2013, Proceedings of the
  23rd International Joint Conference on Artificial Intelligence, Beijing,
  China, August 3-9, 2013}, pages 136--142. {IJCAI/AAAI}, 2013.

\bibitem[\protect\citeauthoryear{Gourv{\`{e}}s \bgroup \em et al.\egroup
  }{2013b}]{gourves2013wine}
Laurent Gourv{\`{e}}s, J{\'{e}}r{\^{o}}me Monnot, and Lydia Tlilane.
\newblock A protocol for cutting matroids like cakes.
\newblock In Yiling Chen and Nicole Immorlica, editors, {\em Web and Internet
  Economics - 9th International Conference, {WINE} 2013, Cambridge, MA, USA,
  December 11-14, 2013, Proceedings}, volume 8289 of {\em Lecture Notes in
  Computer Science}, pages 216--229. Springer, 2013.

\bibitem[\protect\citeauthoryear{Gourv{\`{e}}s \bgroup \em et al.\egroup
  }{2014}]{gourves2014}
Laurent Gourv{\`{e}}s, J{\'{e}}r{\^{o}}me Monnot, and Lydia Tlilane.
\newblock Near fairness in matroids.
\newblock In {\em {ECAI} 2014 - 21st European Conference on Artificial
  Intelligence, Prague, Czech Republic - Including Prestigious Applications of
  Intelligent Systems ({PAIS} 2014), August 18-22.}, pages 393--398, 2014.

\bibitem[\protect\citeauthoryear{Hugo}{1948}]{steinhaus1948}
Steinhaus Hugo.
\newblock The problem of fair division.
\newblock {\em Econometrica}, 16:101--104, 1948.

\bibitem[\protect\citeauthoryear{Kuhn}{1955}]{kuhn1955}
Harold~W. Kuhn.
\newblock The hungarian method for the assignment problem.
\newblock {\em Naval Research Logistics Quarterly}, 2:83--97, 1955.

\bibitem[\protect\citeauthoryear{Kyropoulou \bgroup \em et al.\egroup
  }{2019}]{kyropoulou2019}
Maria Kyropoulou, Warut Suksompong, and Alexandros Voudouris.
\newblock Almost envy-freeness in group resource allocation.
\newblock In {\em Proceedings of the Twenty-Eighth International Joint
  Conference on Artificial Intelligence, {IJCAI-19}}, pages 400--406.
  International Joint Conferences on Artificial Intelligence Organization, 08
  2019.

\bibitem[\protect\citeauthoryear{Lipton \bgroup \em et al.\egroup
  }{2004}]{lipton2004}
Richard~J. Lipton, Evangelos Markakis, Elchanan Mossel, and Amin Saberi.
\newblock On approximately fair allocations of indivisible goods.
\newblock In {\em Proceedings of the 5th {ACM} Conference on Electronic
  Commerce, New York, USA, May 17-20, 2004}, pages 125--131, 2004.

\bibitem[\protect\citeauthoryear{Moulin}{2003}]{moulin2003}
Herv\'{e} Moulin.
\newblock {\em Fair Division and Collective Welfare.}
\newblock MIT Press, 2003.

\bibitem[\protect\citeauthoryear{Pareto}{1897}]{pareto1896}
Vilfredo Pareto.
\newblock Cours d'\'{E}conomie politique.
\newblock {\em Professeur \'{a} l'Universit\'{e} de Lausanne. Vol. I. Pp. 430.
  1896. Vol. II. Pp. 426. 1897. Lausanne: F. Rouge}, 1897.

\bibitem[\protect\citeauthoryear{Plaut and Roughgarden}{2018}]{plaut2018}
Benjamin Plaut and Tim Roughgarden.
\newblock Almost envy-freeness with general valuations.
\newblock In {\em Proceedings of the Twenty-Ninth Annual {ACM-SIAM} Symposium
  on Discrete Algorithms, {SODA} 2018, New Orleans, LA, USA, January 7-10,
  2018}, pages 2584--2603, 2018.

\bibitem[\protect\citeauthoryear{Rawls}{1971}]{rawls1971}
John Rawls.
\newblock {\em A Theory of Justice}.
\newblock Belknap Press of Harvard University Press, Cambridge, Massachussets,
  1 edition, 1971.

\bibitem[\protect\citeauthoryear{Sandomirskiy and
  Segal{-}Halevi}{2019}]{sandomirskiy2019minimal}
Fedor Sandomirskiy and Erel Segal{-}Halevi.
\newblock Fair division with minimal sharing.
\newblock {\em CoRR}, abs/1908.01669, 2019.

\bibitem[\protect\citeauthoryear{Schneckenburger \bgroup \em et al.\egroup
  }{2017}]{schneckenburger2017}
Sebastian Schneckenburger, Britta Dorn, and Ulle Endriss.
\newblock The {A}tkinson inequality index in multiagent resource allocation.
\newblock In Kate Larson, Michael Winikoff, Sanmay Das, and Edmund~H. Durfee,
  editors, {\em Proceedings of the 16th Conference on Autonomous Agents and
  MultiAgent Systems, {AAMAS} 2017, S{\~{a}}o Paulo, Brazil, May 8-12, 2017},
  pages 272--280. {ACM}, 2017.

\bibitem[\protect\citeauthoryear{Sen}{1976}]{sen1976}
Amartya Sen.
\newblock Welfare inequalities and rawlsian axiomatics.
\newblock {\em Theory and Decision}, 7(4):243--262, Oct 1976.

\bibitem[\protect\citeauthoryear{Sen}{1977}]{sen1977}
Amartya Sen.
\newblock Social choice theory: A re-examination.
\newblock {\em Econometrica}, 45(1):53--89, 1977.

\bibitem[\protect\citeauthoryear{Young}{1995}]{young1995}
H.~Peyton Young.
\newblock {\em Equity - in theory and practice}.
\newblock Princeton University Press, 1995.

\end{thebibliography}

\end{document}